%% file: main.tex
\documentclass[journal,draftcls,onecolumn,12pt,twoside]{IEEEtran}
\IEEEoverridecommandlockouts

\input{defs}

\usepackage{ifthen}

\usepackage{mathtools}
\usepackage{multirow}

\mathtoolsset{showonlyrefs=true}

\usepackage{comment}
\newcommand{\ignore}[1]{}

\newcommand{\VersionLength}{short}

\providecommand{\ver}{\ifthenelse{\equal{\VersionLength}{long}}}
\providecommand{\nver}{\ifthenelse{\equal{\VersionLength}{short}}}

\providecommand{\figref}[1]{Fig.~\ref{#1}}
\providecommand{\secref}[1]{Sec.~\ref{#1}}

\providecommand{\ColumnNum}{2}
\newcommand{\col}{\ifthenelse{\equal{\ColumnNum}{1}}}

\begin{document}

\title{Information Velocity of Cascaded \\ Gaussian Channels with Feedback}

\author{Elad Domanovitz, Anatoly Khina, Tal Philosof, and Yuval Kochman
    \thanks{E.~Domanovitz and A.~Khina are with the School of Electrical Engineering, Tel Aviv University, Tel Aviv~6997801, Israel (e-mail: \texttt{\{domanovi,anatolyk\}@eng.tau.ac.il}).}
    \thanks{T.~Philosof is with Samsung Research, Tel Aviv 6492103, Israel (e-mail: \texttt{tal.philosof@gmail.com}).}
    \thanks{Yuval Kochman is with the The Rachel and Salim Benin School of Computer Science and Engineering, the Hebrew University of Jerusalem, Jerusalem 9190401, Israel (email: \texttt{yuvalko@cs.huji.ac.il}).}
}

\maketitle


\begin{abstract}
We consider a line network of nodes, connected by additive white Gaussian noise channels, equipped with local feedback.
We study the velocity at which information spreads over this network.
For transmission of a data packet, we give an explicit positive lower bound on the velocity, for any packet size. Furthermore, we consider streaming, that is, transmission of data packets generated at a given average arrival rate. We show that a positive velocity exists as long as the arrival rate is below the individual Gaussian channel capacity, and provide an explicit lower bound. Our analysis involves applying pulse-amplitude modulation to the data (successively in the streaming case), and using linear mean-squared error estimation at the network nodes. Due 
to the analog linear nature of the scheme, the results extend to any additive noise. For general noise, we derive exponential error-probability bounds. Moreover, for \mbox{(sub-)}Gaussian noise we show a doubly-exponential behavior, which reduces to the celebrated Schalkwijk--Kailath scheme when considering a single node.
Viewing the constellation as an ``analog source'', we also provide bounds on the exponential decay of the mean-squared error of source transmission over the network. 
\end{abstract}


\allowdisplaybreaks

\section{Introduction}
\label{s:intro}

Large-scale, ubiquitous wireless networks are evolving, the Internet of Things being a technology that implements part of the vision.
A central problem of interest in this context, is transmission over a cascade of channels interconnected by relaying nodes. Beyond serving as a building block for more complex networks, such line networks are found, for example,  
in saltatory conduction in the brain, 
where action potentials propagate along myelinated axons with the nodes of Ranvier serving as relays \cite{Waxman-Kocsis-Stys:Axons:Book1995,Girault-Peles:nodes-of-Ranvier:OpNeuroBio2002,Berger-Levy:neural-comm:TIT2010,Suksompong-Berger:integrate-and-fire-capacity:TIT2010}, 
as well as in vehicle platooning, where wireless communication between geographically distant vehicles that travel together in a coordinated fashion is carried over intermediate vehicles which act as relays \mbox{\cite{Heinovski-Dressler:platooning:VNC2018,Nawaz:platooning:TranVehic2019}}. 

From an information-theoretic perspective, 
without delay constraints, assuming that each channel may be used the same number of times, the maximal reliable communication rate is equal to the minimum of the individual channel capacities.  
In a more practical scenario, the end-to-end delay of the network is constrained. Since it is wasteful for a node to wait to decode a long block code, the nodes should opt to apply causal operations to their measurements instead. However, determining the maximal rate of reliable communication 
over such networks, let alone determining the error probability behavior, turns out to be very challenging.

If we fix the message size and number of channels and let the number of time steps grow, decoding with high probability is guaranteed, and one seeks the optimal error exponent. 
Determining the optimal error exponent turned out to be difficult even for the simple case of single-bit transmission over a tandem of binary symmetric channels \cite{Huleihel-Polyanskiy-Shayevitz:real-time-relaying:ISIT2019,Jog-Loh:Real-Time-Relaying:IT2020}, and was eventually proved by Ling and Scarlett~\cite{Ling-Scarlett:tandem-channel:EE:1-bit:TIT2023} to equal to that of a single channel.
They further extended the scope to any finite number of messages in~\cite{Ling-Scarlett:tandem-channel:EE:R=0:TIT2023}.

The behavior of large networks is better expressed by taking the number of channels to grow linearly with the number of time steps. The minimum ratio between the two such that the error probability may be arbitrarily small, was termed \textit{Information Velocity} (IV) by Polyanskiy (see \cite{Huleihel-Polyanskiy-Shayevitz:real-time-relaying:ISIT2019,Rajagopalan-Schulman:FOCS:Real-Time-Networks:STOC1994}) in the single-bit context. The same term was used earlier by Iyer and Vaze \cite{Information-Velocity:Wireless:WiOpt2015} in a related setting of spatial wireless networks.\footnote{Similar concepts also exist in other disciplines, \eg, in physics,
in neuroscience, epidemic spread in networks,
and in marketing and finance.} 

It is not a priori clear whether such a positive IV exists at all for a given network. 
Rajagopalan and Schulman~\cite{Rajagopalan-Schulman:FOCS:Real-Time-Networks:STOC1994}, while considering simulating protocols over binary symmetric channels (BSCs), answered this in the affirmative for a finite number of bits transmitted over BSCs (and hence also for any binary-input output symmetric channels via slicing). 
Improving upon this work, Ling and Scarlett~\cite{Ling-Scarlett:information-velocity::TIT2022} gave tighter upper and lower bounds in the limits of crossover probabilities that go to 0 or $1/2$.

Analyzing the IV of the transmission of a single message does not capture the entire behavior of the network. As the IV is meaningful when both the time and number of relays go to infinity, it might suggest that when another message arrives, the network is still occupied with processing the previous message. Hence, analyzing the IV while transmitting an infinite stream of messages is of interest. For transmission of a stream of messages over a cascade of packet-erasure channels (with instantaneous ACK/NACK feedback), the existence of a positive IV can be derived from stochastic network calculus 
\cite{Fidler2010SurveyOfDetAndStochService, Fidler2015aGuideToTheStichNetCalc,chang2000performance,Fidler2006endtoendProb} (these results also hold for a single packet without feedback). However, only recently, in ~\cite{InformationVelocity:ErasuresWithFeedback:INFOCOM2022},  the explicit IV was derived for streaming over packet-erasure channels with such feedback. This work also derived the IV of transmitting a single packet and determined the explicit error decay rate for velocities below the IV for both single-packet and streaming settings.

Despite these efforts, no explicit expressions for the IV for non-erasure channels are available even for a single bit, let alone for streaming or error probability decay rate guarantees.

\begin{figure}[t]
    \resizebox{\textwidth}{!}{\input{figs/relays_option2.tikz}}
    \caption{Block diagram of the system. $\X{r}{t}$, $\Y{r}{t}$, and $\Z{r}{t}$ are the channel input, output, and noise, respectively, at node $r$ at time $t$. Each $\period$ time steps, a new packet is generated. At every time step, all the nodes decode all the hitherto arrived packets.} 
    \label{fig:relays}
\end{figure}
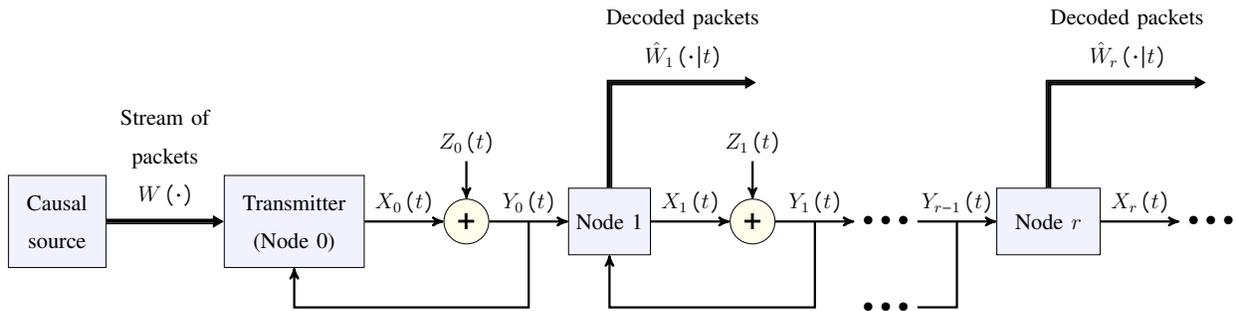

In this work,
we consider a cascade of additive white-noise (say Gaussian) channels with the same signal-to-noise ratio (SNR) $P > 0$, equipped with perfect and local feedback. That is, each node is aware of the measurements of the next node in a causal manner; see Figure~\ref{fig:relays}. For this setting, we derive a lower bound on both the single-packet and the streaming IVs:\footnote{These results are stated for delayed hops, that is, a relay may only use its input in the subsequent time step. In the body of the paper we consider instantenous hops for ease of presentation; the results for both dealyed and instantaneous hops are summarized in Table~\ref{t:summary} in \secref{s:conclusion}.} 
We prove that the single-packet IV is bounded from below~as
\begin{align} 
\label{eq:IV:1packet}
    \IV \geq 1- \Exp{-2 C} ,
\end{align}
where $C \triangleq \half \log(1 + P)$ denotes the channel capacity of an individual additive white-Gaussian noise channel with SNR $P$;
the same bound~\eqref{eq:IV:1packet} has been independently derived by Inovan and Telatar~\cite{inovan-telatar:private-com}.
The streaming IV $\IV(R)$ at data-generation rate $R$ is bounded from below as 
\begin{align}
\label{eq:IV:stream}
    \IV(R) \geq 1 - \Exp{-2(C-R)}
\end{align}
for $R < C$;
in particular, the streaming IV is positive for any $R < C$, and the bound on the streaming IV~\eqref{eq:IV:stream} reduces to that on the single-packet IV~\eqref{eq:IV:1packet} in the limit of $R \to 0$.

Reminiscent of \cite{CausalPM:ISIT2019}, 
our approach is based on the source node translating the incremental message history into a real number. Subsequently, this number is handled by the network in an analog linear manner. Namely, each node keeps the best linear estimator of that number based on its past measurements, as well as the estimate of the next node (known thanks to the feedback), and transmits a scaled version of the difference. Indeed, for a single data packet, the communication between the source node and the first relay reduces to the celebrated Schalkwijk--Kailath scheme for feedback communication over an additive-noise channel \cite{SchalkwijkKailath,Schalkwijk:feedback:finite-BW:TIT1966}, which can be viewed \cite{Gallager-Nakiboglu:SK-via-Elias:TIT2009} as an application of the joint source--channel coding scheme with feedback of Elias \cite{Elias57:JSCC:BW-expansion}.
On the other hand, the first transmission of each of the nodes reduces to scalar amplify-and-forward relaying \cite{ScheinGallager00}.

Treating a representation of a data message as an analog source is a concept that has proved useful in relaying schemes \cite{ScheinGallager00,Laneman-Tse-Worenll:cooperative-diversity:TIT2004,RematchAndForward_Full}. Using that concept, techniques from source coding and joint source--channel coding find their way into digital communication settings. In the context of our work, the estimation at the relays are reminiscent of the Gaussian CEO problem \cite{Yamamoto80,BergerZhangViswanathan:CEO:IT1996,Wagner-Tavilar-Viswanath:Gaussian-lossy-Sepian--Wolf:TITT2008}, and in particular its sequential variant \cite{Draper-Wornell:Sequential-CEO:ISIT2002,Draper-Wornell:SideInfo-sensors:JSC2004,Chen-Zhang-Berger-Wicker:distributed-sensor-networks:Allerton2003}. From the point of view of our relay network, the way that the source node builds its scalar representation of the data stream, is as if an analog source were revealed to it using successive refinement \cite{EquitzCover91,LastrasBerger}. 

Due to the amount of concepts involved, we present our results gradually. We first address a single source to be transmitted, and then streaming. Within each of the above, we start with an analog source and consider the estimation mean-squared error (MSE), and in particular its decay rate, before advancing to data packets and considering error probabilities. For error probabilities, we show that for velocities lower than our achievable bound on the IV, the error probabilities decay at least exponentially fast, while if the noise is assumed to be Gaussian (or, more generally, sub-Gaussian), the error probabilities are shown to decay at least double exponentially, extending the known behavior of the Schalkwijk--Kailath scheme \cite{SchalkwijkKailath,Schalkwijk:feedback:finite-BW:TIT1966} to multiple relays.

The rest of the paper is organized as follows. In Section~\ref{s:model}, we formally define the problem. In Section~\ref{s:1source_sample}, we address the MSE when communicating a single analog source sample. In Section~\ref{s:1packet}, we replace that source sample by a data packet, possibly of unbounded size.
In Section~\ref{s:stream}, we advance to a source sample being successively revealed to the first node, 
and then to a stream of data packets.
Finally, in Section~\ref{s:conclusion} we conclude with a summary of the main results, and discussion of future reserach directions.

\subsection*{Notation}

Denote by $\ints$, $\pints$, $\reals$, $\preals$, and $\nats$ the sets of integer, non-negative integer, real, non-negative real, and natural numbers, respectively.
Throughout, all logarithms and exponents are taken to the natural base.
We make use of small $o$ notation: $f(x) = o_x\lrp{g(x)}$ if $\lim_{x \to \infty} {f(x)}/{g(x)} = 0$; if the argument is clear, we write $o(\cdot)$.
For a scalar $a \in \preals$, we denote
\begin{align} 
\label{eq:bar} 
    \bar{a} = \frac{a}{1+a}. 
\end{align}
We denote time sequences of a signal $a_k$ between times $t_1,t_2 \in \pints$ ($t_1 \leq t_2$) by 
\begin{align}
 \element{a}{k}{t_1:t_2} &\triangleq       
    \begin{bmatrix} 
        a_k(t_1), & a_k(t_1 + 1), & \ldots, & a_k(t_2) 
    \end{bmatrix}.
\end{align}
We denote Markov chains by $X_1 \leftrightarrow X_2 \leftrightarrow X_3$, namely, given $X_2$, $X_1$ is independent of $X_3$.
$\exp$ and $\log$ denote the natural exponential and logarithm functions, respectively. We use the binary entropy and divergence functions: for $0 < p,q < 1$,
\begin{align}
\label{eq:defs:KL-div:entropy}
\begin{aligned}
    h(p) &\triangleq p \log \frac {1}{p} + (1-p) \log \frac{1}{1-p} , 
    \\
    \KL{p}{q} &\triangleq p \log \frac {p}{q} + (1-p) \log \frac{1-p}{1-q}.
\end{aligned}
\end{align}

\section{Problem Statement}
\label{s:model}

\subsection{Communication Model}

The communication model that is considered in this work is depicted in \figref{fig:relays}.

\textit{Causal source.}
Every $\period$ time steps,
a source generates a new message that comprises $\size$ bits, namely, at time $t = \period \tau$ for $\tau \in \pints$, the source generates packet 
\begin{align}
    \packet{\tau} = 
    \begin{bmatrix}
        \bit{\tau \size}, & \bit{\tau \size + 1}, &\ldots, & \bit{(\tau+1) \size - 1}
    \end{bmatrix}
\end{align}
comprising bits. 
The bits of the entire message sequence, $\lrcm{\bit{i}}{i \in \pints}$, are assumed i.i.d.\ uniform. 
Define the average rate of this source, measured in nats per time instant, as
\begin{align}
\label{eq:def:R} 
    R = \frac{\psize}{\period} \log(2) . 
\end{align}

\textit{Channels.} 
The network is composed of a cascade of channels; channel $r \in \pints$ is 
an additive noise channel:
\begin{align} 
\label{eq:channel:output}
    \Y{r}{t} = \X{r}{t} + \Z{r}{t} , 
\end{align}
where $\X{r}{t}$, $\Y{r}{t}$, and $\Z{r}{t}$ are the channel input, output, and noise, respectively, at time $t \in \pints$. The entries of all the noise sequences $\lrc{\Z{r}{t} \middle| r,t \in \pints}$ are i.i.d.\ zero-mean, unit variance, and independent of all channel inputs. All channel inputs are subject to a mean power constraint:
\begin{align}
\label{eq:power-constraint}
    \E{\Xsqr{r}{t}} &\leq P & \forall r,t \in \pints.
\end{align}
The following channel quantities will be useds in the sequel:
\begin{align} \label{eq:C} 
C = \half \log (1+P)
\end{align}
is the Gaussian capacity and (recalling the bar notation \eqref{eq:bar}) 
\begin{align} 
\label{eq:gamma}
\begin{aligned}
    \eta \triangleq (1-\oP)\cdot\Exp{2R} 
  =\Exp{-2(C-R)}.
\end{aligned}
\end{align}
As will be formally defined in the node functions below, perfect feedback is available in each channel, from the channel output to the terminal feeding it, in the subsequent time step.

\textit{Originating transmitter (node 0).}
At each time $t \in \pints$, generates a channel input $\X{0}{t}$ as a function of all the packet history $\packet{0 : \floor{ \frac{t}{T} }}$, and of all past outputs of channel 0, $\Y{0}{0:t-1}$, which are available via feedback. The input is subject to the power constraint \eqref{eq:power-constraint}.

\textit{Node $r$ ($r \in \nats$).} 
At each time $t \in \pints$, generates the channel input $\X{r}{t}$ as a function of all its measurement history\footnote{The current measurement is included, that is, we assume \emph{instantaneous} hops. See Remark~\ref{rem:delayed} in the sequel.} $\Y{r-1}{0:t}$ from its feeding channel (channel $r-1$), and of all its feedback history from the subsequent channel (channel $r$), $\Y{r}{0:t-1}$. The input is subject to the power constraint \eqref{eq:power-constraint}. The relays also produce estimates of the message.\footnote{We consider decoding at a general node $r \in \nats$. This decoding does not affect the creation of subsequent channel inputs in any way. As a special case, one can think of a specific node where the network terminates and information is needed. See Remark~\ref{rem:horizon} in the sequel.} Let $\HB{r}{n}{t}$, $\HB{r}{0:n}{t}$, $\hpacket{r}{n}{t}$, and $\hpacket{r}{0:n}{t}$ denote the estimates of $\bit{n}$, $\bit{0:n}$, $\packet{n}$, and $\packet{0:n}$, respectively, at node $r \in \nats$ at time $t$ based on the measurement history $\Y{r-1}{0:t}$ and feedback history $\Y{r}{0:t-1}$.

\textit{Scheme.}
We refer to the collection of maps of
nodes $0$ to $r \in \nats$ at times $0$ to $t \in \pints$ as an \textit{$(r,t)$-scheme}. 
A \textit{scheme} is defined as a nested set collection of $(r,t)$-schemes with respect to $t \in \pints$ for a fixed $r \in \nats$, and with respect to $r \in \nats$ for a fixed $t \in \pints$,
to wit, an $(r,t)$-scheme equals to the union of all the $(\tr,\tilde{t})$-schemes over $\tr \leq r, \tilde{t} \leq t$.

\subsection{Performance Measures}
{\textit{Error Probability}.}
We define the \textit{bit error probability} for bit $n$ at relay $r$ at time $t$ as 
\begin{align} 
    \label{eq:def:single_bit_pe}
    \pe{r}{n}{t} &\triangleq \PR{\HB{r}{n}{t} \neq \bit{n}}.
\end{align}
We define the \textit{maximal bit error probability} as a function of the detection delay $\Delta$ (the number of time steps elapsed since the generation of the bit) 
\cite{InformationVelocity:ErasuresWithFeedback:INFOCOM2022} as 
\begin{align}
\label{eq:def:max_pe}
    \BLER{r,\Delta}& \triangleq \sup_{\tau\in\pints} \max_{\tau \psize \leq n \leq (\tau+1)\psize - 1} \pe{r}{n} {\tau \period + \Delta} .
\end{align}
The outer maximization (supremum) is carried over all the packets, 
whereas the inner maximization is carried over all the bits inside a packet. $\Delta$ is the elapsed time (delay) after a generated bit is decoded.
Thus, $\BLER{r,\Delta}$ is the worst-case error probability across all the bits recovered $\Delta$ time steps after their generation.

\textit{Achievable streaming velocity.}
A streaming velocity $v \in \preals$ of a source of average rate $R$
is said to be \textit{achievable} if there exists a scheme, such that,
\begin{align}
    \lim_{r \rightarrow \infty} \BLER{r,\floor{\frac{r}{v}}} = 0 ,
\end{align}
namely, $\Delta = \floor{r/v}$ in $P_e$.

\textit{Streaming information velocity.} 
The streaming IV of a source of average rate $R$, denoted by $\IV(R)$,
is defined as the supremum of all achievable streaming velocities of that source.

\begin{figure}[t]
\centering
\includegraphics[scale = 1.2]{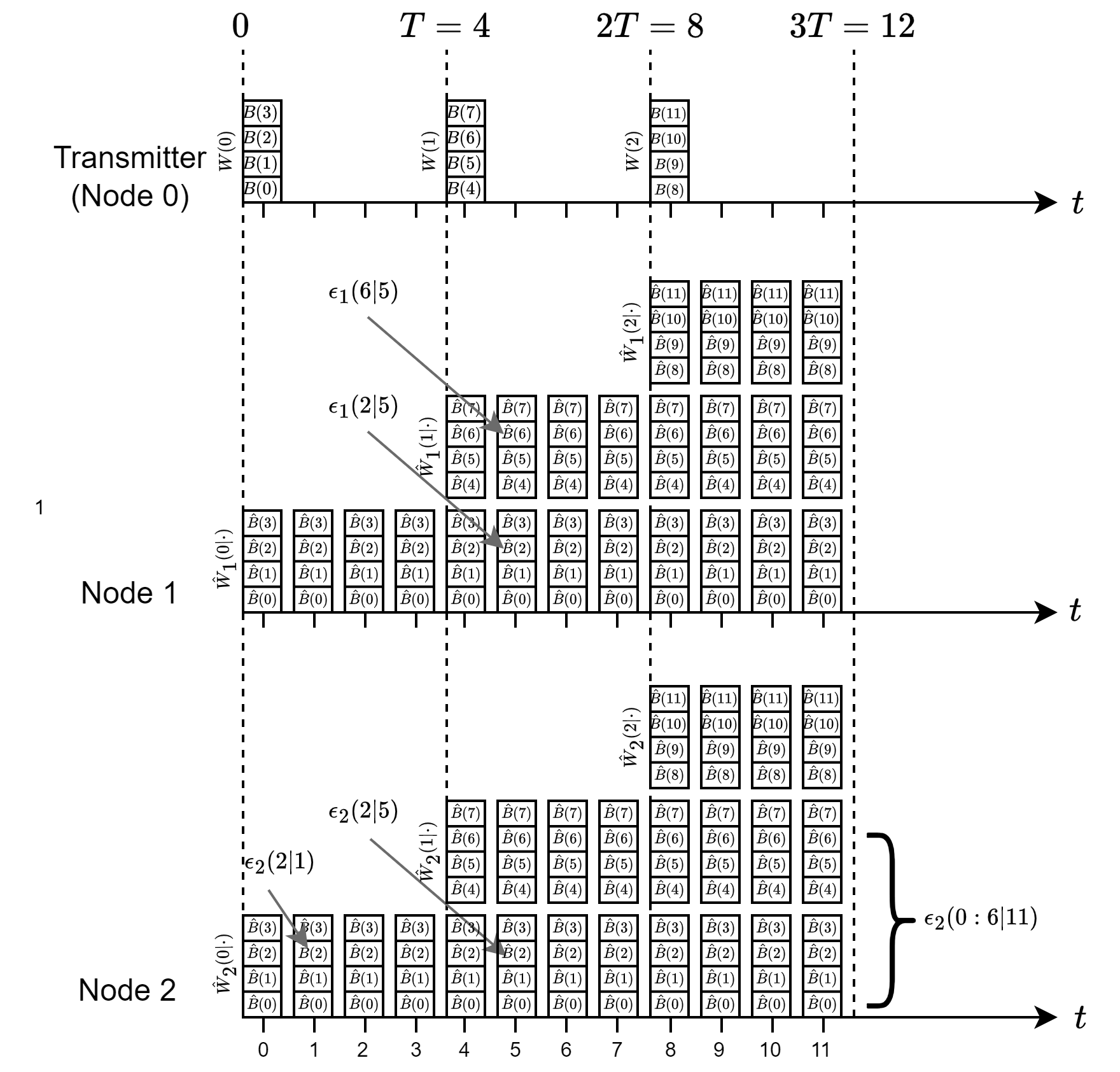}
\caption{Data packet streaming: visualization packet estimations at nodes $r = 0, 1, 2$ across time, for time period $\period=4$, and four bits per packet $\psize=4$. The packets at the transmitter (node $0$) indicate arrival time to transmitter, while the packets at nodes $1$ and $2$ indicate decoded packets.} 
\label{fig:relays-times}
\end{figure}

\figref{fig:relays-times} demonstrates schematically the propagation of packets along the network nodes versus time. The example in \figref{fig:relays-times} consists of a transmitter (node $0$) and two relay nodes (nodes $1$ and $2$). The packet length is $\psize=4$ bits and $T = 4$, meaning that the average rate is $R = \log 2$. The top graph indicates generation times of packets (when they arrive at node $0$). Packets $W(0), W(1), W(2)$ arrive at node $0$ at times $0$, $4$ and $8$, respectively. The middle and bottom graphs depict the estimations of the hitherto generated packet at each time step at nodes 1 and 2, respectively: 
for $t\in [0,3]$, packet $W(0)$ is estimated, namely, $\hat{W}(0|t)$; 
for $t\in[4,7]$, packets $W(0:1)$ are estimated, namely $\hat{W}(0:1|t)$; 
for $t\in[8,11]$, packets $W(0:2)$ are estimated, namely $\hat{W}(0:2|t)$; etc. 

We also mark in \figref{fig:relays-times} a few instances of the error probability, e.g., $\pe{1}{6}{5}$ is the error probability at node $1$ of bit $\bit{6}$ at time $t=5$, and $\pe{2}{0:6}{11}$ is the error probability at node $2$ of bits $\bit{0:6}$ at time $t=11$.

\begin{remark}[average rate] 
The streaming IV is set by the average rate $R$ \eqref{eq:def:R} rather than by the source parameters $\psize$ and $\period$ individually. To see why this is true, notice that the generation time of the bit $n$, which is $\tau \period$ where $\tau$ is the index of the packet containing $n$, equals $ n \log(2) / R$ up to a bounded difference. This difference does not affect the IV as it is absorbed in $\Delta \rightarrow \infty$ in the definition \eqref{eq:def:max_pe}. 
    
        It will become evident that we could use other transmission patterns with the same average rate without changing our results. We keep the periodic pattern for simplicity, noticing that it can approximate any desired positive rate.

\end{remark} 

\begin{remark}[finite horizon]
\label{rem:horizon}
    In this work, we consider a setting of an \textit{infinite} number of cascaded nodes and analyze the velocity at which information can propagate over this network with a decaying error probability. This setting is more stringent than previous IV definitions \cite{Huleihel-Polyanskiy-Shayevitz:real-time-relaying:ISIT2019,InformationVelocity:ErasuresWithFeedback:INFOCOM2022,Ling-Scarlett:information-velocity::TIT2022} in which the number of nodes was a prescribed finite number with a target final recevier, which was then taken to infinity. In particular, in the previously-considered settings, all nodes were designed to attain the best possible error probability at a particular final receiver (``horizon''), whereas in our (``horizon-free'') setting, tension may arise between attaining the best possible performance at different nodes. Clearly, any achievable IV or error probability guarantees carry over to the finite-horizon setting, but not necessarily the other way around.
\end{remark}

\begin{remark}[delayed hops]
\label{rem:delayed}
    The described model above assumes instantaneous hops, namely, 
    node $r$ at time $t$ transmits $\X{r}{t}$ that may dependent 
    on its measurement history $\Y{r-1}{1:t}$ until and \textit{including} time $t$ (and its feedback history $\Y{r}{1:t-1}$).
    The exposition of this scenario is simpler to follow and we therefore present it in detail. That said, we also present results for the parallel scenario of \textit{delayed hops}, in which, node $r$ at time $t$ transmits $\X{r}{t}$ that can depend only on its purely causal history $\Y{r-1}{1:t-1}$ (and its feedback history $\Y{r}{1:t-1}$).
\end{remark}

\section{Detailed Work Overview} 
\label{s:detailed}

The setting described in Section~\ref{s:model} is the ultimate goal of this work. As mentioned in the introduction, we tackle this goal by first considering somewhat simpler settings, namely a single packet (no streaming) and source transmission (the data bits are replaced by a continuous source). Beyond setting the ground for the main result, these settings yield auxiliary results that may be of interest in their own right.

Let $S$ denote a source sample that we want to convey to the nodes with minimal mean-squared error (MSE), and $\HS{r}{t}$ denote its estimate at node $r$ at time $t$. We will analyze the estimation MSE of $\HS{r}{t}$:
\begin{align}
\label{eq:def:MSE}
    \M{r}{t} \triangleq E\left[\left(S-\HS{r}{t}\right)^2\right].
\end{align}

Besides being interesting in and of itself, as a joint source--channel coding (JSCC) problem, the transmission of an analog source over the same network constitutes an essential component in our proposed communication schemes. Specifically, in Section~\ref{s:1source_sample}, the sample $S$ is revealed to the transmitter before transmission begins. 

In Section~\ref{s:1packet}, the error probability of transmitting a single packet over the network is derived. 
Transmitting a single packet may be viewed as a limiting streaming setting in which the packet inter-generation time $T$ is infinite [and the average rate \eqref{eq:def:R} goes to 0]. Therefore, all the definitions of Section~\ref{s:model} carry over to the single-packet transmission setting, except for the maximal bit error probability~\eqref{eq:def:max_pe}. Since the (only) packet comprises $\psize$ bits, $\bit{0:\psize-1}$, that are generated together at time 0, their detection delay $\Delta$ equals simply the elapsed time since the beginning of transmission $t$, the maximal bit error probability reduces, in this case, to 
\begin{align} 
\label{eq:def:max_pe_single}
     \BLER{r,t} &\triangleq \max_{0 \leq n \leq \psize-1 } \pe{r}{n} {t}.
\end{align}
The IV is defined with respect to this error probability. Since our bounds on the single-packet IV do not depend on $\psize$ (and correspond to $R = 0$), we denote the single-packet IV by $\IV$ without an argument.

To bound the maximal error probability of a transmission of a single packet, we will treat the packet
as a source sample and use the bounds on the MSE of transmitting a single source sample. We will subsequently analyze the packet error probability [decoding error probability of $\bit{0:\psize-1}$]:
\begin{align}
    \label{eq:def:PER}
    \pe{r}{0:\psize-1}{t} &= \PR{\HB{r}{0:\psize-1}{t} \neq \bit{0:\psize-1}}.
\end{align}
Clearly, the packet error probability bounds from above
the corresponding individual bit error probabilities since
\begin{align}
\label{eq:FER-vs-BER:1packet}
    \pe{r}{0:\psize-1}{t} &= \PR{\bigcup_{i \in \lrc{0, \ldots, \psize-1}} \lrc{ \HB{r}{i}{t} \neq \bit{i} } }
 \\ &\geq \max_{i\in\{1,\ldots, \psize-1 \}} \PR{\HB{r}{i}{t} \neq \bit{i}}
 \\ &=\max_{i\in\{1,\ldots, \psize-1 \}} \pe{r}{i}{t} .
\end{align}

We then move to analyze streaming settings in Section~\ref{s:stream}.
In Section \ref{s:SourceStreaming}, we consider a streaming JSCC setting in which the sample $S$ is gradually revealed (a la successive refinement) to the transmitter with quality that improves with time and derive bounds on the MSE \eqref{eq:def:MSE} as a function of the node index $r$ and the time $t$.
In \secref{ss:stream:packets}, we use the JSCC streaming solution and results of \secref{s:stream} 
 to derive bounds on the error probability for data-packet streaming by constructing a virtual source comprising the infinite concatenation of the entire data-packet sequence, where the hitherto generated data packets correspond to partial knowledge of the source $S$. 
Similar to \eqref{eq:FER-vs-BER:1packet}, 
to bound the maximal bit error probability~\eqref{eq:def:max_pe}, we bound the more stringent prefix error probability 
\begin{align}
    \label{eq:def:prefixErrorPe}
    \pe{r}{0:n}{t} &= \PR{\HB{r}{0:n}{t} \neq \bit{0:n}} 
\end{align}
for $n \in \nats$, by 
\begin{align}
\label{eq:FER-vs-BER}
\begin{aligned}
     \BLER{r,\Delta} &\triangleq \sup_{\tau\in\pints} \max_{\tau \psize \leq n \leq (\tau+1)\psize - 1} \pe{r}{n} {\tau \period + \Delta} 
  \\ &\leq \sup_{\tau\in\pints} \pe{r}{0:(\tau+1) \psize - 1} {\tau \period + \Delta} .
\end{aligned}
\end{align}


\section{Transmission of a Single Source Sample}
\label{s:1source_sample}

In this section, we consider the problem of transmitting a single zero-mean unit-variance source sample over the network under a minimum MSE criterion. 


We propose the following simple \emph{linear} scheme, which is a natural extension of the single-channel scheme of Elias \cite{Elias57:JSCC:BW-expansion} to our network setting. 

\begin{scheme} 
\label{sc:source_single}
\ 

\noindent
    \textit{Initialization.}
    \begin{itemize}
    \item 
        Since the transmitter (node 0) knows $S$ perfectly, it sets $\HS{0}{t}=S$ for all $t \in \pints$. 
    \item 
        Each node $r \in \nats$ initializes its estimate before transmission begins to the mean: $\HS{r}{-1} = 0$.
    \end{itemize}

\noindent
    \textit{Estimation at node $r \in \nats$.}
    At each time $t \in \pints$, constructs an estimate of $S$:
    \begin{align}
    \label{eq:scheme1_est}
        \HS{r}{t} = \HS{r}{t-1} + \G{r}{t} \Y{r-1}{t} ,
    \end{align}
    where $\G{r}{t}$ is a linear MSE (LMMSE) constant, to be specified in the sequel.

\noindent
    \textit{Transmission by node $r \in \pints$.}
    At each time $t \in \pints$, transmits
    \begin{align}
    \label{eq:scheme1_tx}
        \X{r}{t} = \B{r}{t} \left[\HS{r}{t}-\HS{r+1}{t-1}\right],
    \end{align}
    where $\B{r}{t}$ is a power-normalization constant, to be specified in the sequel.
\end{scheme}

Since node $r \in \nats$ at time $t$ knows $\Y{r+1}{0:t-1}$ via feedback, 
it can construct $\HS{r+1}{t-1}$ which is used to generate the transmit signal~\eqref{eq:scheme1_tx} of node $r$ at time $t$. 

Using \schemeref{sc:source_single}, the estimates satisfy the following properties.
\begin{lem}
\label{lem:SingleSource:LMMSE}
    In Scheme~\ref{sc:source_single}, all channel inputs and outputs and all estimates have zero mean.
    Setting 
    \begin{align} 
    \label{eq:gamma_r}
        \G{r}{t} = \frac{\cov{S}{\X{r-1}{t}}}{1+\var{\X{r-1}{t}}},  
    \end{align}
    we have the following.
    
    \begin{enumerate}
    \item
    \label{itm:LMMSE:all-history}
        $\HS{r}{t}$ is the LMMSE estimate of $S$ from  $\Y{r-1}{0:t}$. 
    \item 
    \label{itm:LMMSE:uncorrelated_outputs}
        The outputs of the same channel at different times are uncorrelated, viz.
        \begin{align}
            \cov{\Y{r}{t}}{\Y{r}{\tau}} &= 0 & t \neq \tau.
        \end{align}
    
    \item 
    \label{itm:LMMSE:cov_Sx}
        The covariance between $S$ and $\X{r}{t}$, for $r,t \in \pints$, equals 
        \begin{align}
            \cov{S}{\X{r}{t}} = \B{r}{t}\cdot \lrs{ \M{r+1}{t-1} - \M{r}{t}}  .
        \end{align}
    \item 
        The variance of the channel input $\X{r}{t}$, for $r,t \in \pints$, equals
        \begin{align}
        \label{eq:var_X}
            \var{\X{r}{t}} = \element{\beta^2}{r}{t} \cdot \lrs{ \M{r+1}{t-1} - \M{r}{t}} .
        \end{align}
    \end{enumerate}
\end{lem}

The proof is based on properties of LMMSE estimation, primarily the orthogonality principle~\cite[Ch.~7-3]{PapoulisBook:4thEd}; it appears in Appendix~\ref{s:1source_sample_proofs}. Notice that the uncorrelatedness of the channel outputs means that indeed each node receives ``novel'' information at each time step, and thus the channels are utilized well. 

Using the properties of \lemref{lem:SingleSource:LMMSE}, and choosing $\B{r}{t}$ to satisfy the power constraint with equality, we find that the constants satisfy,
\begin{subequations}
\label{eq:scheme1_const}
\noeqref{eq:scheme1_est_const}
\begin{align}
    \B{r}{t} &= \sqrt{\frac{P}{\M{r+1}{t-1} - \M{r}{t}} } & \forall r,t \in \pints \,;
\label{eq:scheme1_power_const}
 \\ \G{r}{t}& =  \frac{ \sqrt{P(\M{r}{t-1} - \M{r-1}{t})}}{P+1} & \forall r \in \nats, t \in \pints \,. 
\label{eq:scheme1_est_const}  
\end{align} 
\end{subequations}
Using these values in 
Scheme~\ref{sc:source_single} yields the relation [notice that $\beta_{r-1}(t)\gamma_r(t) = \oP$ for all $r \in \nats$ where $\oP$ is defined according to the notation~\eqref{eq:bar}]:
\begin{align} 
\HS{r}{t} = \oP \cdot \HS{r-1}{t} + (1-\oP) \cdot \HS{r}{t-1} + \G{r}{t} \Z{r-1}{t}, \ \forall r\in\nats, t\in\pints.  \label{eq:HS_recurssion}
\end{align}
Further, the resulting MSE of the scheme is given in the following lemma, whose proof appears in Appendix~\ref{s:1source_sample_proofs}.
\begin{lem}
\label{lem:SingleSource:rec}
    Scheme~\ref{sc:source_single} with the parameters of \eqref{eq:scheme1_const} satisfies the recursion
     \begin{align}
        \M{r}{t} =  \oP\cdot \M{r-1}{t} + (1-\oP)\cdot\M{r}{t-1} \ \forall r\in \nats, t\in\pints,
    \label{eq:SingleSource:rec}
    \end{align}
    with the initial conditions $\M{r}{-1} = 1$ for all $r\in\pints$, and the boundary conditions $\M{0}{t}=0$ for all $t\in\pints$.
Furthermore, the solution of this recursion is 
 \begin{align}
        \M{r}{t} 
        &= \lrr{1-\oP}^{t+1} \sum_{k = 1}^r \binom{t+r-k}{r-k} \oP^{r-k} & \forall r,t\in \nats . 
    \label{eq:EE:1sample:MSE}
\end{align}
\end{lem}


In order to study the asymptotic behavior of \eqref{eq:EE:1sample:MSE}, we fix some velocity $v>0$ such that $t = \floor{ r / v }$, and consider the MSE sequence as a function of the relay index.

\begin{thm} 
\label{thm:SingleSource:EE}
    Let $v > 0$ be some fixed velocity. Then, 
    the MSE of Scheme~\ref{sc:source_single} satisfies
        \begin{align}
            \M{r}{\floor{\frac{r}{v}}}  \leq  \Exp{ - r E_1(v) + o(r)} ,
        \label{eq:SingleSource:MSE}
        \end{align}
    where the $o(r)$ correction is independent of $v$, $E_1(\cdot)$ is defined via $\oP = \frac{P}{1 + P}$~\eqref{eq:bar} as
    \begin{align} 
    \label{eq:E0} 
        E_1(v) \triangleq
        \begin{cases}
            \frac{1}{\ov}\KL{\ov}{\oP}, & v<P; 
         \\ 0, & v \geq P.
        \end{cases}
    \end{align}
\end{thm}
 The proof appears in Appendix~\ref{s:1source_sample_proofs}. It is based on the entropy-based bounds on the binomial coefficients. In fact, It can be shown (although it is not needed for our purposes) that 
 the bound of \eqref{eq:SingleSource:MSE}
 is exponentially tight.

\thmref{thm:SingleSource:EE} implies that, for any $v<P$, the MSE goes to zero with $t$ (and $r \propto t$). 
That is, the source is reconstructed with arbitrarily good precision asymptotically. 
Indeed we can interpret this result as a lower bound on the source ``reconstruction velocity''. 

In the next section, we will use the JSCC results of this section for the problem of transmitting a single data packet.
This is achieved by mapping the data packet to a pulse amplitude modulation (PAM) symbol $S$, applying \ref{sc:source_single} to $S$ by treating the latter as a source samples, and finally decoding the data packet from the resulting LMMSE estimate of $S$. 
This scheme will allow translating the JSCC results of this section to parallel results for data-packet transmission: a lower bound on the single-packet IV of $\IV \geq P$, and an exponential decay rate of $E_1$~\eqref{eq:E0} for velocity $v < P$.

\section{Transmission of a Single Data Packet}
\label{s:1packet}

In this section, we consider the problem of transmitting a single data packet of $\psize$ i.i.d.\ uniform bits $\bit{0:\psize-1}$  over the network.
We bound the block error probability $\pe{r}{0:\psize-1}{t}$ \eqref{eq:def:prefixErrorPe}, which, as explained in \secref{s:detailed}, serves as an upper bound on the worst-case individual bit error probability 
$\BLER{r,t}$ \eqref{eq:def:max_pe_single}. The error-probability bounds yield an achievable single-packet~IV.


To that end, we map the the bits $\bit{0:\psize-1}$, which comprise the data packet, to a PAM constellation point using natural labeling:
\begin{align}  
\label{eq:constellation} 
    S^\psize = \sqrt{3} \sum_{i=0}^{\psize-1} (-1)^{\bit{i}} 2^{-(i+1)}, 
\end{align}  
and apply \schemeref{sc:source_single} by viewing $S^\psize$ as source sample.
Since the bits $\bit{0:\psize-1}$ are uniformly distributed, $S^\psize$ is uniformly distributed over a discrete symmetric finite grid of size $2^\psize$ with the spacing between two adjacent constellation points being 
\begin{align}
\label{eq:min-dist}
    \D{\psize} = \sqrt{3} \cdot 2^{-\psize + 1} ;
\end{align}
in particular, $S^\psize$ has zero mean.
In the limit of an infinite message length $\psize$, $S^\psize$ converges to
    \begin{align}  
    \label{eq:infinite-constellation} 
        S = \lim_{\psize\rightarrow\infty} S^\psize = \sqrt{3} \sum_{i=0}^\infty (-1)^{\bit{i}} 2^{-(i+1)} .
    \end{align}  
    Since the bits $\lrc{\bit{i}}$ are i.i.d.\ uniform, $S$ is uniformly distributed over $[-\sqrt{3},\sqrt{3})$, meaning that it has zero mean and unit variance. For any $\psize$, the variance of $S^\ell$ is smaller than $1$. 
This allows us to construct the following transmission scheme for a single packet, which satisfies the power constraints \eqref{eq:power-constraint}.

\begin{scheme} 
\label{sc:channel_single}
\ 
    \begin{enumerate}
    \item 
        The transmitter (node 0) maps $\bit{0:\psize-1}$ to $S^\psize$ according to \eqref{eq:constellation}. 
    \item 
        All nodes apply Scheme~\ref{sc:source_single} with $S^\psize$ taking the role of the source $S$.
    \item 
    \label{item:channel_single:estimate}
        At each time step $t \in \pints$, each relay (node $r$ for $r \in \nats$) estimates $\bit{0:\psize-1}$ from $\HSfin{\psize}{r}{t}$---the estimate of $S^\psize$ at relay $r$ at time $t$.
    \end{enumerate}
\end{scheme}

\begin{remark}
\label{rem:S:fin:var}
    Since the variance of $S^\psize$ is strictly lower than $1$ (converges to $1$ in the limit of $\psize \to \infty$), 
    \schemeref{sc:channel_single} can be improved by adjusting the power of $S^\psize$ to 1 by multiplying it by a factor that is greater than 1. However, the gain of such an improvement is negligible in the limit of large $t$. 
\end{remark}

We denote the error of $\HSfin{\psize}{r}{t}$ in estimating $S^\psize$ by 
\begin{align}
\label{eq:err:fin}
    \ERRfin{\psize}{r}{t} \triangleq S^\psize - \HSfin{\ell}{r}{t}.
\end{align}
Since $\var{S^\psize} \leq 1$, 
\begin{align}
\label{eq:fin-MSE<inf-MSE}
    \E{\lrc{\ERRfin{\psize}{r}{t}}^2} \leq \M{r}{t} .
\end{align}

\begin{lem} 
\label{lemma:PER:1packet}
    The packet error probability of \schemeref{sc:channel_single} is bounded from above as 
    \begin{align}
        \pe{r}{0:\psize}{t} \leq \frac{1}{3} \cdot 2^{2\psize} \cdot \M{r}{t}
    \end{align}
    with $\M{r}{t}$ of \lemref{lem:SingleSource:rec}.
\end{lem}

\begin{IEEEproof}
    Feeding $\HSfin{\psize}{r}{t}$ to a nearest-neighbor decoder (slicer) of the constellation point $S^\psize$ results in an error probability that is bounded from above as 
    \begin{subequations}
    \label{eq:proof:PER:1packet}
    \begin{align}
        \pe{r}{0:\psize}{t} & \leq \PR{\abs{\ERRfin{\psize}{r}{t}} > \frac{\D{\psize}}{2}}
    \label{eq:proof:PER:1packet:error}
     \\ &\leq \frac{\E{\lrc{\ERRfin{\psize}{r}{t}}^2}}{\D{\psize}^2/4}
    \label{eq:proof:PER:1packet:Chebyshev}
     \\ &\leq \frac{1}{3} 2^{2\psize} \M{r}{t}
    \label{eq:proof:PER:1packet:MSE}
    ,
    \end{align}
    \end{subequations}
    where \eqref{eq:proof:PER:1packet:error} follows from the nearest-neighbor decision rule and the PAM constellation points being distant by $\D{\psize}$, 
    \eqref{eq:proof:PER:1packet:Chebyshev} follows from Chebyshev's inequality, 
    and \eqref{eq:proof:PER:1packet:MSE} follows from \eqref{eq:min-dist} and~\eqref{eq:fin-MSE<inf-MSE}.
\end{IEEEproof}

 Substituting the result of \thmref{thm:SingleSource:EE} in \lemref{lemma:PER:1packet}, and recalling the definition of single-packet IV~\eqref{eq:IV:1packet}, yields immediately the following.

\begin{thm}
\label{thm:1source}
    Let $\psize \in \nats$ be the packet size, however large.
    Then, the single-packet IV \eqref{eq:IV:1packet} is bounded from below as $V \geq P$.

    Moreover, 
    for velocity $v < P$, the achievable prefix-free error probability is bounded 
    as 
    \begin{align}
        \pe{r}{0:\psize}{\floor{\frac{v}{r}}} 
        = \Exp{ - E_1(v) r + o(r) }, 
    \end{align} 
    where $E_1(v)$ was defined in \eqref{eq:E0}.
\end{thm}

For a single channel ($r = 1$) and \textit{Gaussian noise}, 
Schalkwijk and Kailath~\cite{Schalkwijk:feedback:finite-BW:TIT1966} (see also \cite{SchalkwijkKailath,Gallager-Nakiboglu:SK-via-Elias:TIT2009}, \cite[pp.~481--482]{GallagerBook1968}, \cite[Ch.~17.1.1]{ElGamalKimBook}) showed that the error probability of transmitting a single message decays \textit{double exponentially} with $t$:
\begin{align} 
\label{eq:SK-bound:r=1}
    \pe{1}{0:\psize}{t}
    &\leq \exp\big\{-\exp\left\{  2 C t  + o(t) \right\}\big\}.
\end{align}
We next show that the doubly-exponential behavior extends also to our setting of multiple AWGN channels.
To that end, we tighten the bound of \lemref{lemma:Cheby} for AWGN channels as follows.

\begin{lem}
\label{lemma:1packet:double-exp}
    The packet error probability of \schemeref{sc:channel_single} over AWGN channels is bounded 
    as 
    \begin{align}
        \pe{r}{0:\psize}{t} \leq 2 \Exp{- \frac{3}{2^{2\psize + 1} \cdot \M{r}{t}}} 
    \end{align}     
    with $\M{r}{t}$ of \lemref{lem:SingleSource:rec}.
\end{lem}

The proof appears in \appref{app:1packet_AWGN_proofs} and relies on replacing the Chebyshev inequality with respect to the estimation error $\ERRfin{\psize}{r}{t}$ in \eqref{eq:proof:PER:1packet:Chebyshev} with a Chernoff--Hoeffding bound with respect to a \mbox{(sub-)}Gaussian $\ERRfin{\psize}{r}{t}$ in this case.

Substituting the result of \thmref{thm:SingleSource:EE} in \lemref{lemma:1packet:double-exp} yields immediately the following.

\begin{thm}
\label{thm:1sourceGauss}
    Assume AWGN channels and let $\psize \in \nats$ be the packet size, however large.
    Then, for velocity $v < P$, the achievable prefix-free error probability is bounded 
    from above 
    as 
    \begin{align} 
    \label{eq:SK-bound:t=r/v}
        \pe{r}{0:\psize}{\floor{\frac{r}{v}}} 
        = \exp\big\{-\exp\left\{  {E_1(v)} r + o(r) \right\}\big\}, 
    \end{align}
        where $E_1(v)$ was defined in \eqref{eq:E0}.
\end{thm}

\begin{remark}
    The result of \thmref{thm:1sourceGauss} extends to sub-Gaussian noises, 
    since the proof of \lemref{lemma:1packet:double-exp} relies on the sub-Gaussianity of the noise. 
\end{remark}

The result of \thmref{thm:1sourceGauss} may be rewritten in terms of time $t$ as 
\begin{align} 
\label{eq:SK-bound:r=vt}
    \pe{r}{0:\psize}{\floor{\frac{r}{v}}} 
    = \exp\big\{-\exp\left\{  E_1(v) \cdot v t + o(t) \right\}\big\}. 
\end{align}
This expression 
means that the \textit{doubly-exponential} behavior~\eqref{eq:SK-bound:r=1} of the celebrated scheme of Schalkwijk and Kailath~\cite{Schalkwijk:feedback:finite-BW:TIT1966,SchalkwijkKailath} extends also to communication at a fixed velocity across multiple relays as long as the velocity is below $P$.
Indeed, since the error propbability for $r = o(t)$ (and $r = 1$ in particular) can only be better than the limit of Theorem~\ref{thm:1sourceGauss} as $v \rightarrow 0$,
the result of \thmref{thm:1sourceGauss} reduces exactly to \eqref{eq:SK-bound:r=1} in this case, by noticing that $\lim\limits_{v \to 0} E_1(v) v = 2C$ with $C$ of~\eqref{eq:C}. 

\begin{remark}
    Gallager~\cite[pp.~481--482]{GallagerBook1968} (see also \cite{Gallager-Nakiboglu:SK-via-Elias:TIT2009}) rederived the doubly-exponentil decay rate of Schalkwijk and Kailath by introducing a variant of their scheme that sends in the first time step a PAM constellation point, 
    extracts the first Gaussian noise sample via feedback, and then applies over the following time steps the JSCC scheme of Elias~\cite{Elias57:JSCC:BW-expansion} for conveying this Gaussian sample. 
    In the analysis of \schemeref{sc:channel_single} in \appref{app:1packet_AWGN_proofs}, we show that in fact the JSCC scheme can be used from the start to describe the constellation point directly; the analysis of Gallager extends to this case by identifying that the noise is sub-Gaussian with variance proxy that equals the variance. 
\end{remark}

\begin{remark}
    The superexponential behavior of the Schalkwijk--Kailath scheme
    was shown by Gallager and Nakibo\u{g}lu~\cite{Gallager-Nakiboglu:SK-via-Elias:TIT2009} to extend beyond a second-order exponential decay, where the order of the exponential decay is commensurate with the blocklength $n$, namely, a tetration-like decay.
    This behavior seems to extend also to the network setting with a constant velocity $v < P$, but it is beyond the scope of this work.
\end{remark}


The bound of \lemref{lemma:PER:1packet} deteriorate exponentially with an increase in $\psize$. 
We next derive a bound on the prefix error probability $\pe{r}{0:n}{t}$, 
that hold uniformly for all $\psize$ (assuming $n \leq \psize$), and hence serves as a stepping stone for the derivation of similar bounds for streaming in \secref{s:stream}.

\begin{lem} 
\label{lemma:Cheby}
    The prefix error probability of \schemeref{sc:channel_single}, for any $n \leq \psize$, is bounded from above as 
    \begin{align}
        \pe{r}{0:n}{t} \leq \frac{2}{\sqrt{3}} \cdot 2^n \cdot \sqrt{\M{r}{t}}
    \end{align}
    with $\M{r}{t}$ of \lemref{lem:SingleSource:rec}.
\end{lem}

This result is proven in \appref{app:1packet_prefix_proof}.
In conjunction with \thmref{thm:SingleSource:EE}, it allows to prove that any velocity below $P$ is achievable in agreement with \thmref{thm:1source}. However, as
the bound of \lemref{lemma:Cheby} is proportional to the square-root of the bound of of \lemref{lemma:PER:1packet}, it attains only half the error exponent $E_1(v)$.
    This factor two in the exponent is thus the price of decoding in the presence of ``interference'' from an unbounded number of bits that are not decoded.

\section{Streaming} 
\label{s:stream}

In this Section, we consider the packet streaming transmission problem as introduced in \secref{s:model}. Like in single packet transmission, we start with the counterpart JSCC problem, namely, conveying, over the network, a source that is only gradually revealed to the transmitter, and then apply the MSE results to packet streaming.

\subsection{Successively Refined Source} 
\label{s:SourceStreaming}

Suppose that the transmitter node does not have full knowledge of the source at the start ($t=0$), but rather it is given an estimate $\HS{0}{t}$ at time $t$, such that these estimates form a Markov chain $\HS{0}{0} \leftrightarrow \HS{0}{1} \leftrightarrow  \ldots \leftrightarrow S$. Thus, we can think of them as being the reconstructions obtained from a successive refinement scheme feeding the network. In particular, we will assume that these transmitter inputs have MSE
\begin{align} 
\label{eq:initial_MMSE} 
    \M{0}{t-1} &= \exp \{ -2Rt \} & \forall t \in \nats 
\end{align} 
for some $R>0$. We can think of this as a digital message that arrives at $R$ nats per sample (neglecting rounding issues), or as a source that is gradually revealed to the transmitter via a fixed-rate successive refinement scheme.\footnote{This is the MSE that is attained by a sequence of greedy optimal quantizers of rate $R$ (neglecting rounding issues) that are applied to a uniformly distributed source sample. Moreover, for continuous sources, this decay is exponentially optimal in $t$ as it matches the exponent of the distortion--rate function by the Shannon's lower bound and the Gaussian source being the ``least compressable''~\cite[Prob.~10.8 and Th.~10.4.1]{CoverBook2Edition}, \cite[Prob.~3.18, Ch.~3.9]{ElGamalKimBook}.}

The scheme that we use is almost identical to Scheme~\ref{sc:source_single}, except the boundary condition which is not 0 for all $t$ but rather improves according to~\eqref{eq:initial_MMSE}.

\begin{scheme} 
\label{sc:source_streaming}
\ 

\noindent
    \textit{Transmitter initialization.}
    The transmitter (node 0) sets the given $\HS{0}{t}$ for all $t \in \pints$. 
    
The rest of the scheme (initialization of nodes $r \in \nats$, estimation at nodes $r \in \nats$, transmission by nodes $r \in \pints$) is exactly as Scheme~\ref{sc:source_single}.
\end{scheme} 
The power-normalization constant $\B{r}{t}$ and the LMMSE constant $\G{r}{t}$ are set according to~\eqref{eq:scheme1_const}.

\begin{remark}
One may wonder, why the transmitter does not apply preprocessing, estimating the source from its inputs. Indeed, due to the Markov structure, the best estimate of $S$ from $\HS{0}{0:t}$ only depends on $\HS{0}{t}$. There may still be a scalar function, e.g., scaling by a factor, that would improve the estimate; we assume that it is already applied before the estimate is   given to the transmitter.
\end{remark}

In the following theorem, we evaluate the asymptotic behavior of the MSE for a fixed velocity. Namely, we fix some velocity $v>0$, set time $t = \floor{t/v}$, and consider the MSE sequence as a function of relay $r$.

\begin{thm}
\label{thm:SourceStreaming:EE}
    Let $v > 0$ be some fixed velocity. Then, the MSE of \schemeref{sc:source_streaming} satisfies 
    \begin{align}
\M{r}{\floor{\frac{r}{v}}}  \leq \Exp{ - r E_S(v) + o(r) } ,    \end{align}
    where the correction term is independent of $v$, 
    \begin{align} 
        E_S(v) =  
        \begin{cases}
            \frac{1}{1-\eta} \KL{1-\eta}{\oP} + 2R\left(\frac{1}{v}-\frac{\eta}{1-\eta} \right), & 0 \leq v \leq \frac{1-\eta}{\eta};
         \\ \frac{1}{\ov} \KL{\ov}{\oP}, & \frac{1-\eta}{\eta} < v \leq P;
         \\ 0, & P < v;
        \end{cases}
    \label{eq:streaming:EE}
    \end{align}
    and $\eta \triangleq (1-\oP)\cdot\Exp{2R} =\Exp{-2(C-R)}$ as in~\eqref{eq:gamma}.
\end{thm}

The proof appears in Appendix~\ref{s:stream_source_proofs}. 
It is similar to the proofs of Lemma~\ref{lem:SingleSource:rec} and Theorem~\ref{thm:SingleSource:EE}, as the MSE recursion is the same. 
However, the initial conditions for the MSE are different. 
In the first region of \eqref{eq:streaming:EE}, the initial error at the transmitter node is (exponentially) dominant, while, in the second region, the lack of any source knowledge at the relay nodes at $t=0$ is dominant and the exponent is identical to $E_1(v)$ \eqref{eq:E0}. 

The result in \thmref{thm:SourceStreaming:EE} means that, for any $v<P$, the MSE of the source sample goes to zero for an exponentially decaying boundary condition \eqref{eq:initial_MMSE}. That is, the ``reconstruction velocity'' is the same as it were with full a priori source knowledge (Section~\ref{s:1source_sample}).

\begin{figure}[t]
\centering
\includegraphics[scale = 0.5]{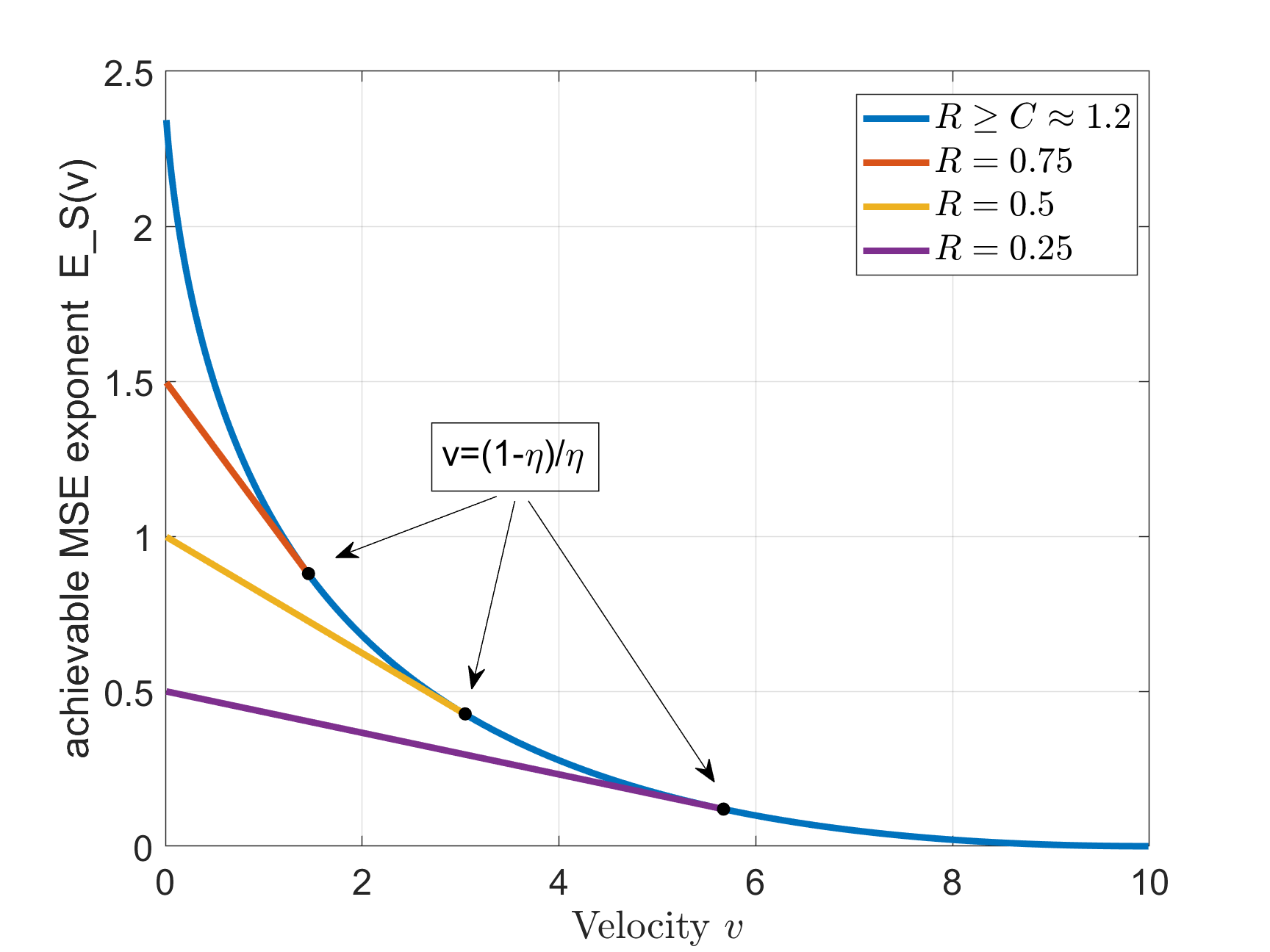}
\caption{Successively refined source: The achievable MSE exponent. The plot depicts the exponent as a function of velocity, at fixed SNR $P=10$, as a function of the refinement rate $R$. The upper curve is $E_1(v)$, which is $E_S(v)$ for any rate above the capacity $C=\log(11)/2\cong 1.2$ nats. Below one sees the first region of the exponent at rates $1$, $0.5$ and $0.1$ nats (from upper to lower curves); in the second region ($\frac{1-\eta}{\eta}\leq v \leq P$) the exponent equals $E_1(v)$. }
\label{fig:MSE-exp}
\end{figure}

    As the rate grows, the exponent $E_S(v)$ becomes the exponent with full source knowledge at the transmitter at $t=0$, $E_1(v)$ \eqref{eq:E0}. This is to be expected, 
    as, in the limit $R \to \infty$, the initial MSE drops immediately. Moreover, for all $R \geq C$, the exponents are already equal, as the first region of \eqref{eq:streaming:EE} is empty in that limit. See Figure~\ref{fig:MSE-exp}. 

    Another way to consider $E_S(v)$, evident in Figure~\ref{fig:MSE-exp}, is the following. Clearly, even at zero velocity, any exponent above $2R$ is not achievable, as this is the exponent at which the transmitter node learns the source. Then, the exponent in the first region is the tangent to $E_1(v)$ at $v=\frac{1-\eta}{\eta}$, originating at $E_S(0) = 2R$.

\subsection{Packet Streaming}
\label{ss:stream:packets}

We now finally reach our target scenario as described in Section~\ref{s:model}. We combine the PAM mapping with the successively-refined source scheme, as follows.
Recall that at time $t=\tau \period$ for $\tau \in \pints$ the packet $\tau$ is made available at the transmitter. Thus, at  time $t = \tau \period$ it has access to bits $\bit{0:\tau\ell-1}$. 
By \eqref{eq:constellation}, the transmitter can then map these bits to the corresponding MAP constellation point
\begin{align} 
\label{eq:constellation_tau}
     S^{\tau\psize} =  \sqrt{3} \sum_{i=0}^{\tau\psize-1} (-1)^{\bit{i}} 2^{-(i+1)}. 
\end{align}
As in Section~\ref{s:1packet}, we define $S$ to be the infinite-constellation limit:
\begin{align} 
   S &=  \lim_{\ell\rightarrow\infty} S^\ell
   = \lim_{\tau\rightarrow\infty} S^{\tau\psize} ;
\label{eq:source_channel}
\end{align}
recall that $S$ has zero mean and unit variance. 
Clearly, the nested constellations have all zero mean, and they form a Markov chain $S^0 \leftrightarrow S^\ell \leftrightarrow S^{2\ell} \leftrightarrow S^{3\ell} \leftrightarrow \ldots \leftrightarrow S$. Furthermore, they form the LMMSE (even MMSE) estimates of $S$ given the available bits, since the resulting estimation error $S-S^{\tau\psize}$ is independent of $\bit{0:\tau\ell-1}$. Thus, we can use Scheme~\ref{sc:source_streaming} for the successively refined source $S$, with $S^{k \ell}$ taking the role of $\HS{0}{\ell}$, as follows.

\begin{scheme} 
\label{sc:channel_stream}
\ 
    \begin{enumerate}
    \item 
        At instants $t=\tau \period$ for $\tau \in \pints$, the transmitter maps $\bit{0:\tau \psize - 1}$ to $S^{\tau \psize}$ according to \eqref{eq:constellation_tau} and updates $\HS{0}{\tau \period} = \HS{0}{\tau \period + 1} = \HS{0}{\tau \period + 2} = \cdots = \HS{0}{(\tau+1) \period - 1}$.
    \item 
        The transmitter and relays apply Scheme~\ref{sc:source_streaming} with respect to the (virtual) source $S$~\eqref{eq:source_channel}.
    \item 
        At each time step $t \in \pints$, each relay (node $r$ for $r \in \nats$) estimates the hitherto generated bits $\bit{0: \lrp{\floor{\frac{t}{T}} + 1} \psize - 1}$
        from $\HS{r}{t}$.
    \end{enumerate}
\end{scheme}

\begin{remark}
\label{rem:packet:stream:MSE<=exp}
    $\HS{0}{t}$ follow an MSE profile in $t$ that is strictly better (smaller) than \eqref{eq:initial_MMSE} for $t \neq \tau \period$ with $\tau \in \pints$, and satisfies \eqref{eq:initial_MMSE} with equality for $t = \tau \period$. Thus, \schemeref{sc:channel_stream} can be improved. However, the gain from such an improvement becomes negligible in the limit of large~$t$.
\end{remark}
    
We can now invoke the MSE of Theorem~\ref{thm:SourceStreaming:EE} and the bound on the error probability of Lemma~\ref{lemma:Cheby}, which holds uniformly for all packet sizes, to obtain the following.

\begin{lem}
\label{lem:connect_exponents}
    In \schemeref{sc:source_streaming} above, for any average rate $R>0$ and for all $v<P$,
    \begin{align}
        \BLER{r,\floor{\frac{r}{v}}} 
        \leq \Exp{- \inf_{t_0 \in \pints }  \lrs{ \frac{t_0 + \Delta}{2v} E_S\left(\frac{v \Delta}{t_0+\Delta}\right)  - t_0 R } + o(r) }, 
    \end{align}
where the worst-bit error probability $\BLER{r,\Delta}$  was defined in \eqref{eq:def:max_pe}, and $E_S(\cdot)$ is given by \eqref{eq:streaming:EE}. 
\end{lem}

The proof appears in Appendix~\ref{s:stream_samples_proofs}. 
It is based upon combining the MSE bound of \thmref{thm:SourceStreaming:EE} with the error-probability bound of Lemma~\ref{lemma:Cheby}. We note that, in \thmref{thm:SourceStreaming:EE}, the velocity is measured from time $0$, 
while in the definition of $\BLER{r,\Delta}$ in \eqref{eq:def:max_pe}, the velocity of a particular packet is measured with respect to the elapsed time since its generation. 
Since each packet is generated at a different time, at a particular node $r$ at a particular time $t$, each packet has a different velocity for the purpose of $\BLER{r,\Delta}$.    
Hence, the key step in the proof of \lemref{lem:connect_exponents} is translating the bound of \thmref{thm:SourceStreaming:EE} using an affine transformation of the velocity.

Having proved this, we are ready to state our main result.

\begin{thm}
\label{thm:main}
    Let $R < C$ be some average rate \eqref{eq:def:R} where $C \triangleq \half \log(1 + P)$ as in~\eqref{eq:C}. 
    Then, the streaming IV is bounded from below as 
    \begin{align} 
        \IV(R) \geq \Exp{2(C-R)} - 1.
    \end{align}
\end{thm}

The proof appears in Appendix~\ref{s:stream_samples_proofs}. It is based upon substituting $E_S(v)$ \eqref{eq:streaming:EE} in the result of Lemma~\ref{lem:connect_exponents}. Noticing that the argument of $E_S(\cdot)$ in \eqref{eq:streaming:EE} is at most $v$, for $v$ below the claimed bound we always take the first region in \eqref{eq:streaming:EE} which gives the desired positive error exponent (it becomes independent of $t_0$ after the substitution, making the minimization redundant). 

\begin{remark}
    For $v > \Exp{2(C-R)} - 1$ and large enough $\Delta$, the argument of $E_S(\cdot)$ in \eqref{eq:streaming:EE} falls inside the second region of \eqref{eq:streaming:EE} and the error exponent becomes negative; we do not provide a proof of this fact, as it is not required for our achievability result. Hence, we believe that the IV bound of \thmref{thm:main} is tight for our \schemeref{sc:channel_stream}.
\end{remark}

\begin{figure}[t]
\centering
    \includegraphics[scale = 0.5]{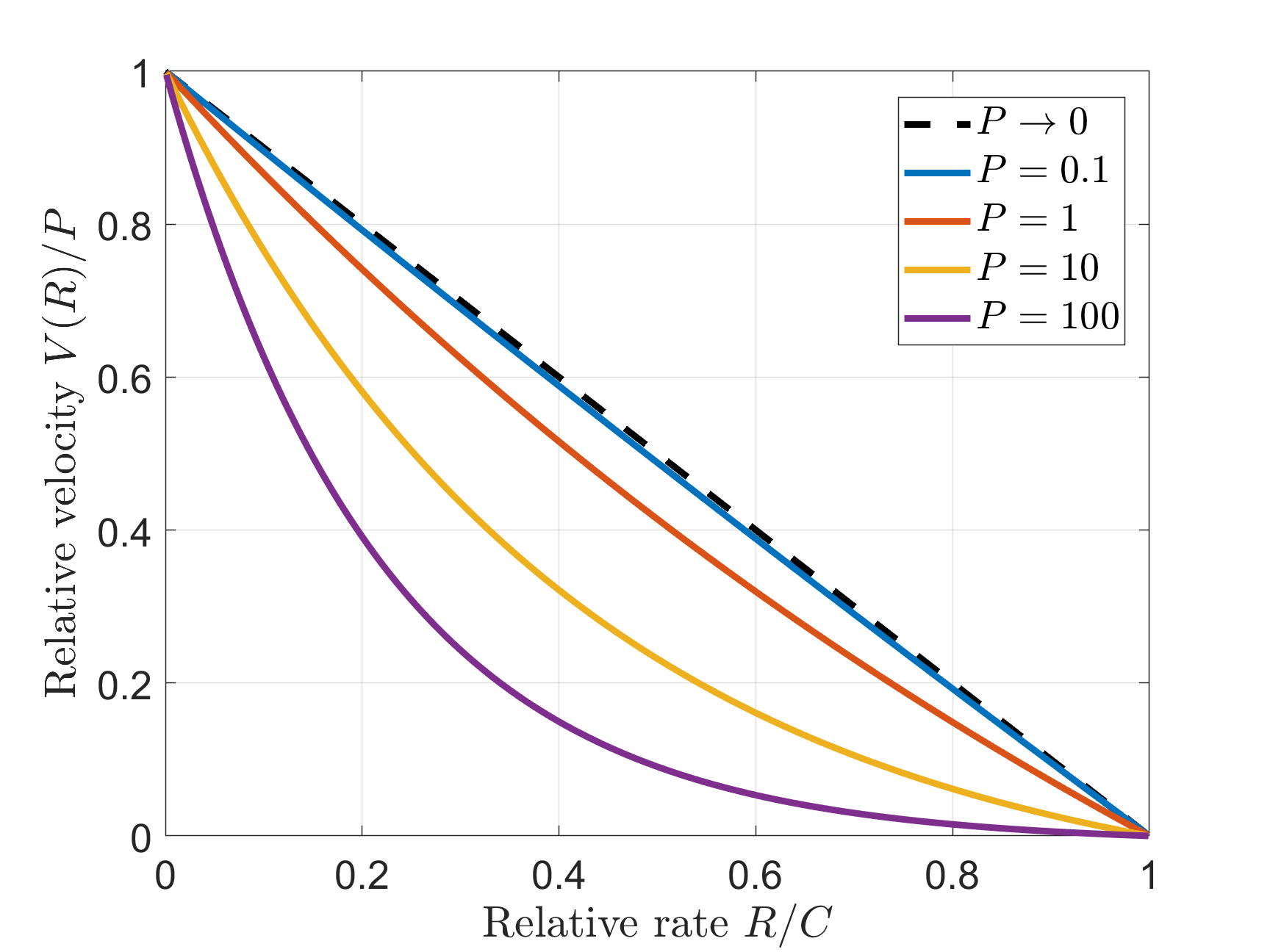}
    \caption{Lower bound on the streaming IV as a function of the rate average rate $R$~\eqref{eq:def:R} for SNRs $P = 0.1, 1, 10, 100$ and the linear limit $P \to 0$. Relative coordinates are used: the lower bound on the streaming IV of \thmref{thm:main} is normalized by the lower bound on the single-packet IV $P$ of \thmref{thm:1source}, whereas the average rate $R$ is normalized by $C$~\eqref{eq:C}.}
\label{fig:relative-V}
\end{figure}

As expected, the IV bound goes to the single-packet IV bound $P$ as the rate goes to zero, and---to zero as the rate goes to $C$. 
It is also worth noting that, for any fixed $P$, the achievable velocity is a convex function of $R$. 
In the low-SNR limit $P\rightarrow 0$, the curve approaches the linear function (See Figure~\ref{fig:relative-V}):
\begin{align} 
    V(R) \cong \left(1-\frac{R}{C}\right) P .
\end{align}
In the high-SNR limit, on the other hand, for any fixed $R$, our bound grows linearly with $P$: 
\begin{align}
    V(R) \cong P \exp\{-2R\}. 
\end{align}


\section{Conclusion and Extensions}
\label{s:conclusion}

We have derived achievable IV and error-probability bounds for both single-packet and streaming communications, and exponential MSE bounds for conveying a source sample, both when it is available upfront and when it is revealed successively to the transmitter.

In Table~\ref{t:summary}, we summarize our main results, and translate them to the setting of delayed hops, which may be more familiar to some of the readers. The translation is immediate, by a linear transformation of the velocity: if the velocity with instantaneous hops is $v$, then with delayed hops it becomes $\ov = v / (1+v)$, and in particular it is at most $1$. We have not included the results on decoding error probability which are longer, but they may be translated in the same way.

\begin{table} 
\centering
\normalsize{
\setlength\tabcolsep{1mm}
\begin{tabular}{|c|c|c|c|c|} 
\hline
     Setting & Reference & Quantity & Instantaneous Hops & Delayed Hops  
\\ \hline \hline  
     \begin{tabular}{c} Single source \\ sample \end{tabular} & Thm. 1 & MSE exponent & 
    $
    \begin{array} {l} 
        \text{For } v \in \left[ 0, P \right):
     \\ \quad \frac{1}{\ov} \KL{\ov}{\oP} 
     \end{array}
    $ 
     &
     $
    \begin{array} {l} 
        \text{For } v \in \left[ 0, \oP \right):
     \\ \quad \frac{1}{v} \KL{v}{\oP} 
     \end{array}
    $
 \\ \hline
    \begin{tabular}{c} Single data \\ packet \end{tabular} & Thm. 2 & \begin{tabular}{c} single-packet \\ IV \end{tabular} & $P$ & $\oP$ 
\\ \hline
    \begin{tabular}{c} Successively-refined \\ source \end{tabular} & Thm. 4 & MSE exponent & 
    $
    \begin{array} {l} 
        \text{For } v \in \left[ 0, \frac{1-\eta}{\eta} \right):
     \\ \quad \frac{1}{1-\eta} \KL{1-\eta}{\oP} 
     \\ \hfill + 2\left(\frac{1}{v}-\frac{\eta}{1-\eta} \right) R
     \\ \text{For } v \in \left[ \frac{1-\eta}{\eta}, P \right):
     \\ \qquad \frac{1}{\ov} \KL{\ov}{\oP}
     \end{array}
    $ 
     &
     $
    \begin{array} {l} 
        \text{For } v \in \left[ 0, \eta \right):
     \\ \quad \frac{1}{1-\eta} \KL{1-\eta}{\oP} 
     \\ \hfill + 2\left(\frac{\ov}{1-\ov} -\frac{\eta}{1-\eta} \right)R
     \\ \text{For } v \in \left[ \eta, \oP \right):
     \\ \qquad \frac{1}{v} \KL{v}{\oP}
     \end{array}
     $
 \\ \hline
    Packet streaming  & Thm. 5 &  Streaming IV  & 
    $\frac{1-\eta}{\eta} $ 
    & $1-\eta$ 
\\ \hline
 \end{tabular}
}
\caption{Summmary of main results, and their translation to delayed hops. Recall that $\oP = P/(1+P)$, $\ov = v / (1+v)$ and $\eta = (1-\oP) \exp\{2R\} = \Exp{-2(C-R)}$.} 
\label{t:summary} 
\end{table}

This work raises many interesting directions for further research.
First, we have only presented achievability results. We believe that our analysis is tight for linear schemes, but is it possible to improve them using a different approach? Specifically, it seems plausible that no positive IV is possible for rates $R>C$, but we are less certain about the IV at lower rates.

Secondly, our analysis presents a rather wide framework that allows for variations. For conveying a source, we considered one that is either known from the start or conveyed at a constant refinement rate; the MSE recursion was the same, and only the initial conditions differed. Other scenarios, including variable-rate refinement and side information of various types in relays along the way may be treated under the same framework. The same holds for data packet settings.

Lastly, 
in this work we have assumed perfect feedback. Such feedback is not practical. In particular, in a line network it does not make sense that, all along a line network, the backward channels enjoy much better conditions than the forward ones. Extending our results to settings with noisy feedback is of major interest. The feedback scheme may be linear, or using modulo-lattice modulation as Ben-Yishai and Shayevitz suggested for a single channel \cite{BenYishaiShayevitz:FeedbackWithWZ:ISIT}. 

\appendices


\renewcommand{\thesectiondis}[2]{\Alph{section}:}

\section{Proofs for Section~\ref{s:1source_sample}}
\label{s:1source_sample_proofs}

We start with two auxiliary lemmas that will be used in the sequel. They refer to an \textit{generalized Scheme~\ref{sc:source_single}}, in which the estimator 
$\HS{r}{t}$ at node $r \in \nats$ at time $t \in \nats$
is taken to be the LMMSE estimator of $S$ from its entire measurement history $\Y{r-1}{0:t}$; we will prove that $\HS{r}{t}$ is of 
the form of~\eqref{eq:scheme1_est} (part~\ref{itm:LMMSE:all-history} of \lemref{lem:SingleSource:LMMSE}). 
We denote  the estimation error~by 
\begin{align} 
    \ERR{r}{\tau} \triangleq S-\HS{r}{\tau}. 
    \label{eq:error} 
\end{align}
\begin{lem} \label{lem:uncorrelated} 
In the generalized Scheme~\ref{sc:source_single}, 
the outputs of the same channel at different times are uncorrelated, viz.\ $\cov{\Y{r}{t}}{\Y{r}{\tau}} = 0$ for $t \neq \tau$.
\end{lem}

\quad\textit{Proof:}
Assume without loss of generality that $0 \leq \tau < t$. Then, 
\begin{subequations} 
\label{eq:cor_YY}
\begin{align}
    &\!
    \cov{\Y{r}{t}}{\Y{r}{\tau}} 
    \stackrel{(a)}= \cov{\X{r}{t}+\Z{r}{t}}{\Y{r}{\tau}}
    \stackrel{(b)}= \cov{\X{r}{t}}{\Y{r}{\tau}}
 \\ &\ \ \stackrel{(c)}=  \cov{ \beta_{r}(t) \lrs{\HS{r}{t} - \HS{r+1}{t-1}} }{\Y{r}\tau}
 \stackrel{(d)}=  \cov{ \beta_{r}(t) \lrs{\ERR{r+1}{t-1} - \ERR{r}{t}} }{\Y{r}\tau}
\nonumber
 \\ &\ \ \stackrel{(e)}=  \beta_{r}(t) \lrs{ \cov{ \ERR{r+1}{t-1} }{\Y{r}\tau} - \cov{\ERR{r}{t} }{\Y{r}\tau} }
    ,
\end{align}
\end{subequations} 
where 
$(a)$ follows from \eqref{eq:channel:output},
$(b)$ holds since $\Z{r}{t}$ is independent of $\Y{r}{0:t-1}$ and hence uncorrelated with its entries,
$(c)$ follows from \eqref{eq:scheme1_tx},
$(d)$ follows from adding and subtracting $S$ and \eqref{eq:error},
and 
$(e)$ follows from the bilinearity of the covariance.

Now, to complete the proof, we prove that both of the resulting elements equal zero as follows. 
\begin{enumerate}
\item
    Since $\ERR{r+1}{t-1}$ is the LMMSE estimation error given $\Y{r}{0:t-1}$, 
    by the orthogonality principle \cite[Ch.~7-3]{PapoulisBook:4thEd},
    $\cov{\ERR{r+1}{t-1}}{\Y{r}{0:t-1}} = 0$. In particular, since $\tau \leq t-1$, 
    $\cov{\ERR{r+1}{t-1}}{\Y{r}{\tau}} = 0$.   
\item 
\label{itm:E(r)-orthog-Y(r-1)}
    $\Y{r}{\tau}$ is a linear combination of $\Y{r-1}{0:\tau}$ and $Z_r(0:\tau)$. 
    By the orthogonality principle,
    $\cov{\ERR{r}{t}}{\Y{r-1}{0:t}}$. $\ERR{r}{t}$ is independent of (and hence uncorrelated with) the noise process $Z_r$ of the subsequent channel. Since $\tau \leq t-1$, $\cov{\ERR{r}{t}}{\Y{r}{\tau}} = 0$.
    \hfill\IEEEQED
\end{enumerate}
%

\begin{lem} 
\label{lem:mean_identity} 
    In the generalized Scheme~\ref{sc:source_single}, 
    \begin{align} 
        \cov{\ERR{r}{t}}{\ERR{r+1}{t-1}} 
        &= \var{\ERR{r}{t}} &\forall r,t \in \pints.
    \end{align}
\end{lem}

\begin{IEEEproof}
    $\begin{aligned}[t]
        \cov{\ERR{r}{t}}{\ERR{r+1}{t-1}} 
        &\stackrel{(a)}= \cov{\ERR{r}{t}}{S - \HS{r+1}{t-1}}
        \stackrel{(b)}= \cov{\ERR{r}{t}}{S}
     \\ &\stackrel{(c)}= \cov{\ERR{r}{t}}{\HS{r}{t} + \ERR{r}{t}}
        \stackrel{(d)}= \var{\ERR{r}{t}}
    \end{aligned}$
    
    \noindent
    where 
    $(a)$ and $(c)$ follow from \eqref{eq:error};
    and 
    $(b)$ and $c$ follow from $\cov{\ERR{r}{t}}{\Y{r-1}{0:t}}$
    and $\cov{\ERR{r}{t}}{\Y{r-1}{0:t}} = 0$, respectively,
    which were proved at the end of the proof of \lemref{lem:uncorrelated}, by noting that $\HS{r+1}{t-1}$ and $\HS{r}{t}$ are linear combinations of $\Y{r}{0:t-1}$
    and $\Y{r-1}{t}$, respectively.
\end{IEEEproof}

\quad\textit{Proof of \lemref{lem:SingleSource:LMMSE}:}

\begin{enumerate}
\item 
    Consider $\HS{r}{t}$ for $r \in \nats$ of the generalized Scheme~\ref{sc:source_single}. 
    By Lemma~\ref{lem:uncorrelated}, the entries of $\Y{r-1}{0:t}$ are uncorrelated, meaning that the LMMSE estimator $\HS{r}{t}$ equals 
    \begin{align} 
    \label{eq:vector-LMMSE-->scalar-LMMSE}
        \HS{r}{t} = \sum_{\tau =0}^{t} \frac{\cov{S}{\Y{r-1}{\tau}}}{\var{\Y{r-1}{\tau}}} \cdot \Y{r-1}{\tau} 
        = \sum_{\tau = 0}^{t} \frac{\cov{S}{\X{r-1}{\tau}}}{1 + \var{\X{r-1}{\tau}}} \cdot \Y{r-1}{\tau}
        , 
    \end{align}
    where the second equality follows from \eqref{eq:channel:output}, the independence of $S$ and the process $Y_{r-1}$, and $\var{\Z{r-1}{\tau}} = 1$ for all $\tau$. 
    Eq.~\eqref{eq:vector-LMMSE-->scalar-LMMSE} suggests the following recursive form:
    \begin{align} 
    \label{eq:est-S:recursive}
        \HS{r}{t} 
        = \HS{r}{t-1} + \frac{\cov{S}{\X{r-1}{t}}}{1 + \var{\X{r-1}{t}}} \cdot \Y{r-1}{\tau}
        = \HS{r}{t-1} + \G{r}{t} \Y{r-1}{\tau}
        , 
    \end{align}
    where the second equality follows from \eqref{eq:gamma_r}.
    
    This proves that Scheme~\ref{sc:source_single} and
    the generalized Scheme~\ref{sc:source_single} are in fact identical.
\item Follows immediately from the proof of part~1 of the lemma and from Lemma~\ref{lem:uncorrelated}. 
\item
    By \eqref{eq:scheme1_tx}, 
    $\cov{S}{\X{r}{t}}= \B{r}{t} \lrc{\cov{S}{\HS{r}{t}} - \cov{S}{\HS{r+1}{t-1}}}$. 
    Substituting
    \begin{align}
        \cov{S}{\HS{r}{t}} 
        &\stackrel{(a)}= \cov{S}{S} - \cov{S}{\ERR{r}{t}}
        \stackrel{(b)}= \var{S} - \cov{\HS{r}{t} + \ERR{r}{t}}{\ERR{r}{t}}
    \nonumber
     \\ &\stackrel{(c)}= \var{S} - \var{\ERR{r}{t}}
        \stackrel{(d)}= \var{S} - \M{r}{t} , 
    \end{align}
    the proof follows, 
    where 
    $(a)$ and $(b)$ hold by the definition of $\ERR{r}{t}$~\eqref{eq:error},
    $(c)$ follows from the orthogonality principle, 
    and $(d)$ hold by the definition of $\M{r}{t}$~\eqref{eq:def:MSE} and ~\eqref{eq:error}.
    
\item
    $\begin{aligned}[t]
        \var{\X{r}{t}}
        &\stackrel{(a)}= \Bsqr{r}{t} \cov{\HS{r}{t} - \HS{r+1}{t-1}}{\HS{r}{t} - \HS{r+1}{t-1}}
     \\ &\stackrel{(b)}= \Bsqr{r}{t} \cov{\ERR{r}{t} - \ERR{r+1}{t-1}}{\ERR{r}{t} - \ERR{r+1}{t-1}}
     \\ &\stackrel{(c)}= \Bsqr{r}{t} \lrs{ \var{\ERR{r}{t}} + \var{\ERR{r+1}{t-1}} - 2 \cov{\ERR{r}{t}}{\ERR{r+1}{t-1}} }
     \\ &\stackrel{(d)}= \Bsqr{r}{t} \lrs{ \var{\ERR{r}{t}} - \var{\ERR{r+1}{t-1}} }
     ,
    \end{aligned}$\\[-.3\baselineskip]

    where
    $(a)$ follows from \eqref{eq:scheme1_tx},
    $(b)$ follows from \eqref{eq:error},
    and 
    $(d)$ follows from \lemref{lem:mean_identity}.
    \hfill \IEEEQED
\end{enumerate}

\begin{IEEEproof} [Proof of \lemref{lem:SingleSource:rec}]
    $\M{r}{t} \stackrel{(i)}= 1 - \sum_{\tau=1}^t \frac{\covSQR{S}{\Y{r-1}{\tau}}}{\var{\Y{r-1}{\tau}}} \stackrel{(ii)}= 1 - \sum_{\tau=1}^t \frac{\covSQR{S}{\X{r-1}{\tau}}}{P+1}$,\\
    where $(i)$ follows 
    from the LMMSE expression~\cite[Eq.~(7-85)]{PapoulisBook:4thEd} and parts \ref{itm:LMMSE:all-history} and \ref{itm:LMMSE:uncorrelated_outputs} of Lemma~\ref{lem:SingleSource:LMMSE}; 
    and $(ii)$ follows from 
    \eqref{eq:channel:output}, the independence of $S$ and the unit-variance noise process $Z_{r-1}$, 
    and by choosing $\lrcm{\B{r-1}{\tau}}{\tau \in \pints}$ such that
    $\lrcm{\X{r-1}{\tau}}{\tau \in \pints}$ satisfies the power constraint \eqref{eq:power-constraint} with equality.
    This can be rewritten recursively as 
    \begin{align}
        \M{r}{t-1} - \M{r}{t}  &=   \frac{\covSQR{S}{\X{r-1}{t}}}{P+1}
        \stackrel{(a)}=  \Bsqr{r-1}{t} \cdot \frac{[\M{r}{t-1} - \M{r-1}{t}]^2}{P+1}
    \nonumber
     \\ &\stackrel{(b)}=  \oP\cdot\lrs{\M{r}{t-1} - \M{r-1}{t}}
        ,
    \end{align}
    where 
    $(a)$ follows from part~\ref{itm:LMMSE:cov_Sx} of Lemma~\ref{lem:SingleSource:LMMSE},
    and
    $(b)$ follows from \eqref{eq:scheme1_power_const}. This proves \eqref{eq:SingleSource:rec}.

    We next solve the recursion \eqref{eq:SingleSource:rec} explicitly.
    Due to the linearity of the recursion, $\M{r}{t}$ equals the sum of the effects of all initial conditions $\M{k}{-1} = 1$ for $1 \leq k \leq r$ (the zero boundary conditions $\M{0}{t} = 0$ have zero effect).
    Consider all the trajectories from $(k,-1)$ to $(r,t)$, where at each step along the trajectory either the time index increases by $1$---corresponding to a multiplication by $1-\oP$, or the relay index increases by 1---corresponding to a multiplication  the MSE by $\oP$. There are $t+1$ time steps and $r-k$ relay steps. Multiplying by the combinatorial expression for the number of trajectories, we have that the effect of a single condition is
    \begin{align}
        \binom{t+r-k}{r-k} (1-\oP)^{t+1} \oP^{r-k}. 
    \end{align} 
    Eq.~\eqref{eq:EE:1sample:MSE} then follows by summing over all relevant $k$.
\end{IEEEproof}

\begin{IEEEproof}[Proof of \thmref{thm:SingleSource:EE}]
    The claim is trivial
    for $v \geq P$  
    since $ \M{r}{t} \leq \var{S} = 1$. 
    Thus, we assume $v < P$ and bound the LMMSE at node $r \in \nats$ at time $t \in \pints$ as follows.
    \begin{align}
        \M{r}{t} 
     &= (1-\oP)\sum_{k = 1}^r \binom{t+r-k}{r-k}\Exp{-(t+r-k) \lrs{ \KL{\frac{r-k}{t+r-k}}{\oP} + \h{\frac{t}{t+r-k}} }}
     \\ &
        \leq \Exp{ - \min_{k \in \lrc{1, 2, \ldots, r}} (t + r - k) \KL{\frac{r-k}{t+r-k}}{\oP} + \log r} ,
     \end{align}
    where the equality follows from \eqref{eq:EE:1sample:MSE} from \lemref{lem:SingleSource:rec} and the definitions of $\h{\cdot}$ and $\KL{\cdot}{\cdot}$~\eqref{eq:defs:KL-div:entropy},
    and the inequality follows from bounding the average by the maximum and from
    ~\cite[Ch.~11.1]{CoverBook2Edition} 
    \begin{align}
    \label{eq:n_choose_k}
        {n \choose i} 
        \leq \Exp{-n \, \h{\frac{i}{n}}} 
        .
    \end{align}
    
    By taking $t=\floor{r/v}$ as in the theorem statement and defining
    $\lambda \triangleq \frac{r -k}{t}$, 
    we have
\begin{subequations} 
\begin{align}
    \M{r}{\floor{\frac{r}{v}}}  &
    \leq \Exp{ - \floor{\frac{r}{v}} \inf_{\lambda \in (0, v)} (1 + \lambda) \KL{\frac{\lambda}{1 + \lambda}}{\oP} + \log r} 
\label{eq:EE:1sample:cont}
 \\ &
   \leq \Exp{ - \left(\frac{r}{v} -1\right) (1+v) \KL{\ov}{\oP} + \log r } 
\label{eq:EE:1sample:cont2}
 \\ &
    \leq \Exp{ - \frac{r}{\ov} \KL{\ov}{\oP} - (1+P) \log{\oP} + \log r } ,
\label{eq:EE:1sample:explicit}
\end{align}
\end{subequations}
    where the second inequality
    holds since the expression in the infimum is monotonically decreasing in $\lambda$ (for $v<P$) and by using $\ov = \frac{v}{1+v}$~\eqref{eq:bar} and $\floor{r/v} \geq r/v - 1$. The last inequality holds since $\KL{\ov}{\oP} \leq -\log{\oP}$ for all $v<P$.
\end{IEEEproof}



\section{Proof of \lemref{lemma:1packet:double-exp}}{
\label{app:1packet_AWGN_proofs}

{To prove \lemref{lemma:1packet:double-exp}, we first introduce auxiliary definitions and results.

We will make use of the following notion of sub-Gaussianity as it appears in \cite[Ch.~2.1]{Wainwright:high-dim-stat:Book2019}.\footnote{Other definitions that are equivalent up to constants exist, e.g., in \cite[Ch.~2.5]{Vershynin:high-dim-prob:Book2018}.} 

\begin{defn}
\label{def:subGauss}
    An RV $X$ 
    with mean $\mu = \E{X}$ is sub-Gaussian with 
    \textit{variance proxy}
    $\SIGsqr{X} > 0$
    if 
    \begin{align}
        \E{\Exp{\lambda(X-\mu)}} &\leq \Exp{\frac{\SIGsqr{X}\lambda^2}{2}}
      & \forall \lambda \in \reals .
    \end{align}
\end{defn}
Clearly, if $X$ is sub-Gaussian with 
variance proxy $\SIGsqr{X}$, it is also sub-Gaussian with any variance proxy greater than $\SIGsqr{X}$.
The following well-known properties follow immediately from \defnref{def:subGauss} and Chernoff's bound.}

\begin{proper}
\label{proper:subGauss}
\ 

    \begin{enumerate}
    \item 
        A Gaussian RV $X$ with variance $\sigma^2$ is sub-Gaussian with $\SIGsqr{X} = \sigma^2$. 
    \item 
        A Rademacher RV (uniformly distributed over $\lrc{\pm 1}$) $X$ is sub-Gaussian with $\SIGsqr{X} = 1$. 
%
    \item 
        Let $X_1, \ldots, X_n$ be independent sub-Gaussian RVs with variance proxies $\sigma^2_1,  \ldots, \sigma^2_n$, respectively.
        Let $a_1, \ldots, a_n \in \reals$ be some constants.
        Then, $\sum\limits_{i=1}^n a_i X_i$ is sub-Gaussian with
        ~$\sum\limits_{i=1}^n a_i^2 \sigma_i^2$.
    \item 
        \textit{Chernoff--Hoeffding bound:} Let $X$ be a zero-mean sub-Gaussian RV with variance proxy $\SIGsqr{X} > 0$, and let $a$ be some positive constant. Then, 
        \begin{align}
            \PR{\abs{X} \geq a} \leq 2 \Exp{-\frac{a^2}{2 \SIGsqr{X}}} .
        \end{align}
    \end{enumerate}
\end{proper}

We now return to \schemeref{sc:channel_single} and prove the following property on 
$\ERRfin{\psize}{r}{t}$ of \eqref{eq:err:fin}.

\begin{lem}
\label{lem:error_is_subGauss}
    $\ERRfin{\psize}{r}{t}$ is sub-Gaussian with variance proxy that satisfies $\SIGsqr{\ERRfin{\psize}{r}{t}} \leq \M{r}{t}$.
\end{lem}

\begin{IEEEproof}
    The inputs to the overall system are the noise samples $\lrcm{\Z{r}{t}}{r,t \in \pints}$
    and $S^\psize$, all of which are independent. The noise samples are Gaussian, 
    whereas $S^\psize$ equals a linear combination of independent Rademacher RVs by \eqref{eq:constellation} (which are also independent of the Gaussian noise samples), and hence also $\ERRfin{\psize}{r}{t}$.
    By parts 1 and 2 of Property~\ref{proper:subGauss}, for all the elements in the linear combination, the variance proxy equals the variance, while by part 3 the same holds for $\ERRfin{\psize}{r}{t}$. Thus, 
    $\ERRfin{\psize}{r}{t}$ 
    is sub-Gaussian with 
    \begin{align}
        \SIGsqr{ \ERRfin{\psize}{r}{t} } 
        = \E{ \lrc{\ERRfin{\psize}{r}{t}}^2 }  \leq \M{r}{t} ,
    \end{align}
    where the inequality follows from \eqref{eq:fin-MSE<inf-MSE}.
\end{IEEEproof}

\vspace{.5\baselineskip}
\begin{IEEEproof}[Proof of \lemref{lemma:1packet:double-exp}]
    $\begin{aligned}[t]
        \pe{r}{0:\psize}{t} 
        & \stackrel{(a)}\leq \PR{\abs{\ERRfin{\psize}{r}{t}} > \D{\psize}/2}
        \stackrel{(b)}\leq 2 \Exp{- \frac{\D{\psize}^2/4}{2 \M{r}{t}}}
     \\ &\stackrel{(c)}\leq 2 \Exp{- \frac{3}{2^{2\psize + 1} \cdot \M{r}{t}}},
    \end{aligned}$

    \vspace{.5\baselineskip}
    \noindent
    where ${(a)}$ follows from \eqref{eq:proof:PER:1packet:error},
    $(b)$ follows from the Chernoff--Hoeffding bound (recall Property~\ref{proper:subGauss}) 
    and \lemref{lem:error_is_subGauss},
    and 
    $(c)$ follows from \eqref{eq:min-dist}.
\end{IEEEproof}


\section{Proof of \lemref{lemma:Cheby}}
\label{app:1packet_prefix_proof}


    To make the analysis devoid of $\psize$, 
    we will analyze the error probability of $S^n$ in the infinite-constellation limit $S$ \eqref{eq:infinite-constellation}. 
    Due to the linearity of all the components of the scheme until the final decoding of the bits, one can always attain the performance of decoding $S^n$ from a finite-constellation $\psize \in \nats$ by adding, to the term that is proportional to $S^\psize$ in the decoding error, 
    a uniform noise $U^\psize$ over $[-\D{\psize}/2, \D{\psize}/2)$ that is independent of $S^\psize$ with the same coefficient, 
    such that $S$ is uniform over $[-\sqrt{3},\sqrt{3})$: 
    \begin{align}  
    \label{eq:infinite-constellation:ell} 
        S = S^\psize + U^\psize, 
    \end{align}  
    with $S^\psize$ as in \eqref{eq:constellation} and 
    \begin{align}
    \label{eq:U}
        U^\psize = \sqrt{3} \sum_{i=\psize}^\infty (-1)^{\bit{i}} 2^{-(i+1)};
    \end{align}
    this will be be made precise in the sequel in \eqref{eq:hSell-->hS}.
    
    Similarly, the decomposition \eqref{eq:infinite-constellation:ell} applies for any $n$:
    \begin{align}
    \label{eq:S=Sn+Un}
        S = S^n + U^n, 
    \end{align}
    where $U^n$ is given in \eqref{eq:U} with $\psize$ replaced by $n$, 
    it is independent of $S^n$ and is uniformly distributed over $[-\D{n}/2,\D{n}/2)$ with 
    \begin{align} 
    \label{eq:min-dist:inf} 
        \D{n} = \sqrt{3} \cdot 2^{-n + 1} 
    \end{align} 
    being the spacing between two constellation levels of $S^n$, in agreement with \eqref{eq:min-dist}.
    
    Now notice that in Scheme~\ref{sc:source_single}, the estimate of $S^\psize$ at node $r \in \nats$ at time $t \in \pints$, is given by
    \begin{align}
    \label{eq:hS:fin}
        \HSfin{\psize}{r}{t} = \A{r}{t} \cdot S^\psize + \Zeff{r}{t},
    \end{align}
    where $\Zeff{r}{t}$ is a combination of channel noises, independent of $S^\psize$, and by the properties of MMSE estimation (recall Lemma~\ref{lem:SingleSource:LMMSE})
    $\A{r}{t} \in (0, 1)$. 
    By \eqref{eq:infinite-constellation:ell}, adding $\A{r}{t} U^\psize$ to $\HSfin{\psize}{r}{t}$ results~in 
    \begin{align}
    \label{eq:hSell-->hS}
        \HS{r}{t} = \A{r}{t} \cdot S + \Zeff{r}{t}.
    \end{align}


        To bound the error probability of decoding $S^n$ at node $r$ at time $t$, 
        we will analyze a suboptimal decoder that 
        \begin{enumerate}
        \item 
            generates the estimate $\HSfin{\psize}{r}{t}$ of step \ref{item:channel_single:estimate} of \schemeref{sc:channel_single};
        \item 
            generates a randomly generated noise $U^\psize$ as in \eqref{eq:U}, which is uniformly distributed over $[-\D{\psize}/2, \D{\psize}/2)$ and is independent of $S^\psize$ and $\lrc{\Z{r}{t}}$;
        \item
            adds $\A{r}{t} U^\psize$ to $\HSfin{\psize}{r}{t}$ [recall \eqref{eq:hS:fin}] to generate $\HS{r}{t}$ of \eqref{eq:hSell-->hS};
        \item 
            decodes $S^n$ from $\HS{r}{t}$ using a nearest-neighbor decoder (slicer) of  $S^n$.
        \end{enumerate}
        The prefix error probability of $S^n$ is bounded from above as follows.
        \begin{subequations}
        \label{eq:proof:1packet:PER:inf-packet}
        \begin{align}
            \pe{r}{0:\psize}{t} 
            &\leq \PR{\abs{S^n - \HS{r}{t}} \geq \D{n}/2}
        \label{eq:proof:1packet:PER:inf-packet:NN}
         \\ &= \PR{\abs{(1-\A{r}{t}) S^n - \A{r}{t} U^n - \Zeff{r}{t}} \geq \D{n}/2}
        \label{eq:proof:1packet:PER:inf-packet:explicit-err}
         \\ &= \int_{-\D{n}/2}^{\D{n}/2} \frac{1}{\D{n}} \PR{\abs{(1-\A{r}{t}) S^n - \A{r}{t} u - \Zeff{r}{t}} \geq \D{n}/2} du
        \label{eq:proof:1packet:PER:inf-packet:U-unif}
         \\ &\leq \frac{1}{\D{n}} \int_{-\D{n}/2}^{\D{n}/2} \PR{\abs{(1-\A{r}{t}) S^n - \Zeff{r}{t}} \geq \D{n}/2 - \A{r}{t} \cdot \abs{u}} du
        \label{eq:proof:1packet:PER:inf-packet:triang-ineq}
            ,
        \end{align}
        \end{subequations}
        where 
        \eqref{eq:proof:1packet:PER:inf-packet:NN} follows from the nearest-neighbor decision rule and since the values that $S^n$ can take are distant by at least $\D{n}$,
        \eqref{eq:proof:1packet:PER:inf-packet:explicit-err} follows from \eqref{eq:S=Sn+Un} and \eqref{eq:hSell-->hS},
        \eqref{eq:proof:1packet:PER:inf-packet:U-unif} holds since $U^n$ is uniformly distributed over $[-\D{n}/2, \D{n}/2)$ and is independent of $\lrp{\Zeff{r}{t}, S^n}$,
        \eqref{eq:proof:1packet:PER:inf-packet:triang-ineq} follows from the triangle inequality by noting that $0 < \A{r}{t} < 1$.

        Now define $\ERRfin{n}{r}{t} = (1-\A{r}{t}) S^n - \Zeff{r}{t}$.
        Note that $\ERRfin{n}{r}{t}$ can be interpreted as the estimation error $S^n - \HSfin{n}{r}{t}$ as in \eqref{eq:err:fin}, 
        where $\HSfin{n}{r}{t}$ is the estimate of $S^n$ if the packet were of size $n$.
        Hence, $\pe{r}{0:\psize}{t}$ can be further bounded as follows.
        \begin{subequations}
        \begin{align}
            \pe{r}{0:\psize}{t} 
             &\stackrel{(a)}= \frac{2}{\D{n}} \int_{[1-\A{r}{t}] \D{n}/2}^{\D{n}/2} \PR{\abs{\ERRfin{n}{r}{t}} \geq x } dx
            \stackrel{(b)}\leq \frac{2}{\D{n}} \int_0^\infty \PR{\abs{\ERRfin{n}{r}{t}} \geq x } dx
         \\ &\stackrel{(c)}\leq \frac{2}{\D{n}} \int_0^\infty \min \lrc{1, \frac{\M{r}{t}}{x^2}} dx
            \stackrel{(d)}\leq \frac{2}{\D{n}} \lrs{ \int_0^{\sqrt{\M{r}{t}}} 1 dx + \int_{\sqrt{\M{r}{t}}}^\infty \frac{\M{r}{t}}{x^2} dx }
        \nonumber
         \\ &\stackrel{(e)}= \frac{4}{\D{n}} \sqrt{\M{r}{t}}
            \stackrel{(f)}= \frac{2}{\sqrt{3}} \cdot 2^n \cdot \sqrt{\M{r}{t}} \,,
        \end{align}
        \end{subequations}
        where 
        $(a)$ follows from \eqref{eq:proof:1packet:PER:inf-packet} by noting that the integrand is an even function, 
        by the definition of $\ERRfin{n}{r}{t}$, and by integration by substitution with $x = \frac{\D{n}}{2} - \alpha u$;
        $(b)$ holds since the integrand is non-negative (and $0 < \A{r}{t} < 1$);
        $(c)$ holds by Chebyshev's inequality and 
        \eqref{eq:fin-MSE<inf-MSE} (the same argument applies here for $n$ in lieu of $\psize$), and since the probability is trivially bounded from above by 1;
        and $(f)$ follows from \eqref{eq:min-dist:inf}.
        \hfill \IEEEQED


\section{Proof of \thmref{thm:SourceStreaming:EE}}
\label{s:stream_source_proofs}

%

When analyzing Scheme~\ref{sc:source_streaming}, all of the analysis that we applied to Scheme~\ref{sc:source_single} up to the recursion relation \eqref{eq:SingleSource:rec} in Lemma~\ref{lem:SingleSource:rec} remains valid. In particular, with the  parameters $\B{r}{t}$ and $\G{r}{t}$ 
of \eqref{eq:scheme1_const}, $\{\HS{r}{t}\}$ are the LMMSE estimators from all available channel outputs (which are uncorrelated), and the power constraint is always satisfied with equality. To see this, 
one may go through the proofs and see that the assumption on the nature of $\{\HS{0}{t}\}$ wasn never used.

Consequently, the MSE recursion of \eqref{eq:SingleSource:rec} remains valid for Scheme~\ref{sc:source_streaming}, 
albeit with different boundary conditions~\eqref{eq:initial_MMSE} [and the same initial conditions $\M{r}{-1} = 1$, for all $r \in \pints$].
 
We first provide an explicit expression for the MSE, parallel to \eqref{eq:EE:1sample:MSE}:
\begin{align} \label{eq:SourceStreaming:MSE} \M{r}{t} =   \MSE^\text{I}_{r}(t) + \MSE^\text{II}_{r}(t), \end{align} 
where
\begin{align}
\MSE^\text{I}_r(t) & = \lrr{1- \oP}^{t+1} \sum_{k = 1}^r \binom{t+r-k}{r-k} \oP^{r-k} , \\
\MSE^\text{II}_r(t) 
& = \oP^{r} \sum_{s=0}^{t-1} \Exp{-2R (t-s)} \binom{r+s-1}{s}  \lrp{1-\oP}^{s}
.
\end{align}
This MSE follows by summing the contributions of all the initial conditions, as done when proving \eqref{eq:EE:1sample:MSE}. $\MSE^\text{I}_r(t)$ is the sum of contributions of the initial conditions $\M{r}{-1} = 1$ for $r > 0$, and indeed it equals the MSE of \eqref{eq:EE:1sample:MSE}.  $\MSE^\text{II}_r(t)$ is the sum of contributions of the boundary conditions $\M{0}{t} = 1$ for $t > 0$, which can be derived in a dual way to the first contribution, substituting the roles as follows: $r\rightarrow t$, $t\rightarrow r-1$, $\oP \rightarrow 1-\oP$, taking into account the different boundary conditions.

Now we move to the second stage, asymptotics. 
Since $\MSE^\text{I}_r(t)$ is identical to single-source MSE of \eqref{eq:EE:1sample:MSE}, by \thmref{thm:SingleSource:EE}, we have that
\begin{align}
    \MSE^\mathrm{I}_{r}{\left(\floor{\frac{r}{t}}\right)} &\leq \Exp{ - r E_1(v) +o(r)} ,
\label{eq:EE:MSE1}
\end{align}
where $E_1(v)$ is defined in \eqref{eq:E0}. 

For $\MSE^\text{II}_r(t)$, we proceed in a similar manner to the proof of Theorem~\ref{thm:SingleSource:EE}:
\begin{subequations}
\label{eq:EE:SourceSample}
\begin{align}
    \MSE^\text{II}_r(t) &=  \oP\, \Exp{-2Rt} \sum_{s = 0}^{t-1} \binom{r+s-1}{s} 
 \\ & \qquad \cdot \Exp{-(r+s-1) \lrs{ \KL{\frac{s}{r+s-1}}{1-\oP} + \h{\frac{s}{r+s-1}} }+2Rs}
\label{eq:EE:SourceSample:IT} 
 \\ & 
    \leq \Exp{-2Rt} \sum_{s = 0}^{t-1}  \Exp{ - \lrs{(s + r -1 ) \KL{\frac{s}{s+r-1}}{1-\oP}-2Rs }} .
\label{eq:EE:SourceSample:n_choose_k} 
\end{align}
\end{subequations}
 Denote the summands as
$a_s = \Exp{ - (r-1) \tE \left(\frac{s}{r-1}\right)}$ for
 \begin{align} \label{eq:tilde_E}
     \tE(\delta) = (1+\delta)\KL{\bar{\delta}} {1-\oP} - 2 R \delta.
 \end{align}
Proceeding as in the proof of Theorem~\ref{thm:SingleSource:EE} would result in a correction term $o(t)$ that grows with $t$. Instead, we derived a bound with a correction term $o(r)$ that does not depend on $t$ and would prove more favorable.
To that end, notice that $\tE(\delta)$ is a convex function, with derivative
\begin{align} 
    \tE'(\delta) = \log\left( \frac{\bar{\delta}}{1-\oP}\right) - 2R = \log\left(\frac{\bar\delta}{\eta} \right),   
\end{align}
which attains its (unconstrained) minimum at
    $\delta^* = \frac{\eta}{1-\eta}$ where $\eta$ was defined in \eqref{eq:gamma}. 

Now, set $s^*$ to be the index of the maximal element $a_s$ in the sum. We have two cases:
\begin{enumerate}
\item If
$(r-1)\delta^* \geq t-1$ the maximum is attained on the edge $s^*=t-1$, 
and
\begin{align}  
\label{eq:sum:as:internal-max}
    \sum_{s = 0}^{t-1}  a_s &\leq t a_{s^*} \leq [(r-1)\delta^* + 1] a_{s^*}. 
\end{align}
\item The maximum is attained at n internal point such that $\abs{s^*-(r-1)\delta^*} \leq 1$, and 
 \begin{align}
   \sum_{s = 0}^{t-1} a_s \leq \sum_{s = 0}^\infty a_s = \sum_{s = 0}^{2s^*+1} a_s + \sum_{s = 2(s^*+1)}^\infty a_s.  
 \end{align}
 The first term in \eqref{eq:sum:as:internal-max} is bounded from above as 
 \begin{align}
    \sum_{s = 0}^{2s^*+1} a_s 
    \leq 2(s^*+1) a_{s^*} \leq 2 \lrs{(r-1)\delta^* + 2} a_{s^*}.      
 \end{align}
 To bound the second term, notice that, by convexity,
 \begin{align}
 \tE \left(\frac{s}{r-1}\right) \geq \tE \left(\frac{2(s^*+1)}{r-1}\right) + \frac{s-2(s^*+1)}{r-1} \tE' \left(   \frac{2(s^*+1)}{r-1} \right) ,
 \end{align}
and, by invoking convexity and $\abs{s^*-(r-1)\delta^*} \leq 1$, we can further bound the derivative~as
\begin{align} \tE' \left(   \frac{2(s^*+1)}{r-1} \right) \geq \tE'(2\delta^*) = \log \frac{2}{1+\eta}. \end{align}
Consequently, all the elements in the second sum in \eqref{eq:sum:as:internal-max} are bounded from above as
\begin{align} a_s \leq a_{2(s^*+1)} \left(\frac{1+\eta}{2}\right)^{s-2(s^*+1)}, \end{align}
and the infinite sum itself is bounded as
\begin{align} 
    \sum_{s = 2(s^*+1)}^\infty a_s \leq \sum_{s = 2(s^*+1)}^\infty a_{2(s^*+1)} \left(\frac{1+\eta}{2}\right)^{s - 2(s^*+1)} \leq \frac{2}{1-\eta} a_{2s^*}
    ,
\end{align}
where the geometric sum is finite since $\eta<1$ by definition~\eqref{eq:gamma}. Since $a_{2s^*}$ decays exponentially with $r$ with an exponent that is greater than that of $a_s^*$, $a_{2s^*} = o_r \lrp{a_{s^*}}$.
\end{enumerate}
We conclude that there exists a \textit{linear} function $c(r)$ that does not depend on $v$ (for both cases), such that
\begin{align}
    \sum_{s = 0}^{t-1} a_s \leq c(r) a_{s^*}.
\end{align}
It follows that
\begin{align}
\label{eq:EE:SourceSample:explicit}  
    \MSE^\text{II}_r\left(\floor{\frac{r}{v}}\right)
    &\leq \Exp{ - r E_2(v) + o(r)}, 
 \\ E_2(v) &\triangleq v \inf_{\delta \in \left(0,\frac{1}{v}\right]} \tE(\delta) - 2R, 
\label{eq:SourceStreaming:EE2}
\end{align} 
where $\tilde E(\cdot)$ was defined in \eqref{eq:tilde_E} and the redundancy term $o(r)$ is independent of $v$.

Since the infimum is $\delta^*$ for $v < \eta/(1-\eta)$ and the edge $1/v$ otherwise, 
$E_2(v) = E_S(v)$ for all $v$. 
Since $E_2(v)\leq E_1(v)$ for all $v$, we obtain the desired result:
\begin{align} 
\M{r}{\floor{\frac{r}{v}}} &\leq  \Exp{ - r E_1(v) + o(r)} + \Exp{ - r E_2(v) + o(r)} \\
&= \Exp{ - r \min(E_1(v),E_2(v)) + o(r)} = \Exp{ - r E_S(v) + o(r)} ,
\end{align}
where the correction term $o(r)$ is independent of $v$ as required.
\hfill \IEEEQED


\section{Proofs for Section~\ref{ss:stream:packets}}
\label{s:stream_samples_proofs}

\begin{IEEEproof}[Proof of Lemma~\ref{lem:connect_exponents}]
    Consider first the boundary condition. By the construction of \schemeref{sc:channel_stream}, 
    \begin{align}
    \noeqref{eq:stream:ch:boundary}
    \label{eq:stream:ch:boundary}
        \!
        \M{0}{t} &= \E{[S-\HS{0}{t}]^2} 
        = \E{\left[S-S^{\psize\floor{\frac{t}{\period}}}\right]^2} 
        = \E{\lrp{U^{\ell \floor{\frac{t}{T}}}}^2} 
        = \frac{ \left( \D{\psize\floor{\frac{t}{\period}}}\right)^2}{12} 
        = 2^{-2\psize\floor{\frac{t}{\period}}} ,\
        \quad 
    \end{align}
    where $U^n$ was defined in \eqref{eq:U} for any $n \in \nats$ and is uniformly distributed over $[-\D{n}/2, \D{n}/2)$, and $\D{n}$ was defined in \eqref{eq:min-dist:inf}. 
    Recalling the rate definition \eqref{eq:def:R}, 
    we have
    \begin{align}  
        \M{0}{t} \leq \exp\{-2Rt\}.
    \end{align}
    Thus, 
    indeed the MSE at node $r$ and time $t$ is at most $\M{r}{t}$ of Scheme~\ref{sc:source_streaming} with source refinement rate $R$ that equals our packet streaming rate $R$. This MSE $\M{r}{t}$, in turn, is given by \eqref{eq:SourceStreaming:MSE}, 
    and is bounded as in   Theorem~\ref{thm:SourceStreaming:EE}.
    
    We now bound the maximal bit error probability \eqref{eq:def:max_pe} as follows.
    \begin{subequations}
    \label{eq:proof:packet:stream:BER:non-asymp}
   \begin{align}
    \BLER{r,\Delta}
   &\leq \sup_{\tau\in\pints} \eps_r (0:(\tau+1)\psize - 1| \tau \period + \Delta) \label{eq:not_ugly_a}
     \\ 
        &\leq \sup_{\tau\in\pints}  \frac{2}{\sqrt{3}} \cdot 2^{(\tau+1)\psize - 1} \cdot \sqrt{\M{r}{\tau \period + \Delta}}
        ,\quad \label{eq:not_ugly_b}
    \end{align}
        \end{subequations}
    where \eqref{eq:not_ugly_a} follows from \eqref{eq:FER-vs-BER}; 
    and \eqref{eq:not_ugly_b} follows from \lemref{lemma:Cheby}
    with $\M{r}{t}$ of Scheme~\ref{sc:source_streaming}, by nothing that its proof in \appref{app:1packet_prefix_proof} holds for any boundary conditions $\HS{0}{t}$.

    To obtain the required asymptotic bound, we 
    substitute $r=\ceil{v \Delta}$ in
    \eqref{eq:proof:packet:stream:BER:non-asymp}: 
    \begin{subequations}
    \label{eq:proof:stream:packet:BER:asymp}
    \begin{align}
        \BLER{\ceil{v \Delta},\Delta} &\leq \frac{2}{\sqrt{3}} \sup_{\tau\in\pints}  2^{(\tau+1)\psize - 1} \cdot \sqrt{\M{\ceil{v \Delta}}{\tau \period + \Delta}} 
    \label{eq:proof:stream:packet:BER:asymp:non-asymp}
     \\ & = \sup_{\tau\in\pints} \exp\left\{ R \tau T -\frac{\tau \period + \Delta}{2v} E_S\left( \frac{v \Delta}{\tau T + \Delta} \right)  + o(\Delta) \right\} 
    \label{eq:proof:stream:packet:BER:asymp:subs}
     \\ & \leq \sup_{t_0\in\pints} \exp\left\{ R t_0 -\frac{t_0 + \Delta}{2v} E_S\left( \frac{v \Delta}{t_0 + \Delta} \right)  + o(\Delta) \right\},
    \label{eq:proof:stream:packet:BER:asymp:all-nats}
    \end{align}
    \end{subequations}
    where 
    \eqref{eq:proof:stream:packet:BER:asymp:non-asymp} follows from \eqref{eq:proof:packet:stream:BER:non-asymp}; 
    \eqref{eq:proof:stream:packet:BER:asymp:subs}
    follows from \eqref{eq:def:R}, \thmref{thm:SourceStreaming:EE},
    and noting that $E_s(\cdot)$ is a smooth function;
    and 
    \eqref{eq:proof:stream:packet:BER:asymp:all-nats} holds since $\period \pints \subset \pints$.
    The required result follows by taking the $o(\Delta)$ correction term out of the optimization. This holds true since the correction term is independent of the argument of $E_S(\cdot)$, as shown in Theorem~\ref{thm:SourceStreaming:EE}.
\end{IEEEproof}


\begin{IEEEproof}[Proof of Theorem~\ref{thm:main}]
Recalling \eqref{eq:gamma}, we bound $\BLER{v,\floor{r/v}}$ for 
\begin{align}
\label{eq:proof:stream:channel:first-region}
    v \leq \Exp{2(C-R)}- 1 = \frac{1 - \eta}{\eta}
\end{align}
as follows.
\begin{subequations}
\label{eq:proof:IV:stream:channel:BER}
\begin{align} 
    \BLER{r,\floor{\frac{r}{v}}} 
    &\leq \Exp{- \inf_{t_0 \in \pints }  \lrs{ \frac{t_0 + \Delta}{2v} E_S\left(\frac{v \Delta}{t_0+\Delta}\right)  - t_0 R } + o(r) }
\label{eq:proof:IV:stream:channel:BER:from_lemma}
 \\ &= \Exp{ - \lrs{\frac{1}{1-\eta} \left( \frac{\KL{1-\eta}{\oP}}{2} - \eta R \right) + R } \Delta+ o(r)}
    ,
\label{eq:proof:IV:stream:channel:BER:rearrange}
\end{align}
\end{subequations}
where
\eqref{eq:proof:IV:stream:channel:BER:from_lemma} follows from Lemma~\ref{lem:connect_exponents},
and 
\eqref{eq:proof:IV:stream:channel:BER:rearrange} holds for $v$ that satisfies \eqref{eq:proof:stream:channel:first-region} by noting that the argument of $E_S(\cdot)$ is at most $v$ in this case.

Now note that
the exponent in \eqref{eq:proof:IV:stream:channel:BER:rearrange} at $v = \frac{1 - \eta}{\eta}$ is positive:
\begin{align} 
    \Bigl. \frac{1}{1-\eta} \left( \frac{\KL{1-\eta}{\oP}}{2} - \eta R \right) + R \Bigr|_{v = \frac{1-\eta}{\eta}} = \frac{\KL{1-\eta}{\oP}}{2(1-\eta)} > 0   
    .
\end{align}
Since $\BLER{r,\floor{r/v}}$ is monotonically increasing with $v$, $\lim_{\Delta \to \infty} \BLER{r,\floor{r/v}} = 0$ for all $v$ that satisfy \eqref{eq:proof:stream:channel:first-region}. Thus all these velocities are achievable, and the proof is concluded.
\end{IEEEproof}

\bibliographystyle{IEEEtran}
\bibliography{toly}

\end{document}

%% file: defs.tex
\usepackage[tbtags]{amsmath} 
\usepackage{amssymb}  
\usepackage{verbatim} 
\usepackage{amsxtra}  
\usepackage{comment}

\usepackage{mathabx}

\usepackage{enumerate}

\usepackage{amsfonts}
\usepackage{color}

\usepackage{subcaption}
\usepackage{wasysym}

\usepackage{cite}

\usepackage{amsthm}

\newtheorem{thm}{Theorem}
\newtheorem{lem}{Lemma}
\newtheorem{proper}{Property}

\theoremstyle{definition}
\newtheorem{defn}{Definition}
\newtheorem*{defn*}{Definition}
\newtheorem{scheme}{Scheme}
\newtheorem*{scheme*}{Scheme}

\theoremstyle{remark}
\newtheorem{remark}{Remark}

\usepackage{fdsymbol}

\usepackage{tikz}

\usetikzlibrary{shapes,arrows,decorations.markings,positioning}

\tikzstyle{plant} = [draw, fill=red!5, rounded rectangle, 
    minimum height=3em]
\tikzstyle{block} = [draw, fill=blue!5, rectangle, 
    minimum height=3em]
\tikzstyle{tap} = [draw, fill=olive!10, rectangle, minimum height=3em]
\tikzstyle{sum} = [draw, fill=yellow!10, circle, node distance=25mm]
\tikzstyle{pinstyle} = [pin edge={to-,thick,black}]

\tikzstyle{BitPipe} = [thick, decoration={markings,mark=at position
   1 with {\arrow[semithick]{triangle 60}}},
   double distance=1.4pt, shorten >= 5.5pt,
   preaction = {decorate},
   postaction = {draw,line width=1.4pt, black, shorten >= 4.5pt}]

\usetikzlibrary{patterns,calc}

\usetikzlibrary{chains,shapes.multipart}
\tikzstyle{FIFO} = [rectangle split, rectangle split parts=3, draw, rectangle split horizontal,minimum height=3em,text height=1.5em,text depth=1em,on chain,inner ysep=0pt]

\usetikzlibrary{fit,backgrounds}

\tikzstyle{coord} = [coordinate]
\tikzstyle{gain} = [draw, fill=red!5, regular polygon, regular polygon sides=3, shape border rotate=-90]

\tikzset{mwe/.style={minimum size=0.5em}}

\usetikzlibrary{fit,backgrounds}

\usepackage{hyperref}

\usepackage{epsfig, graphicx, psfrag}

\usepackage[mathcal]{euscript}
\usepackage[cal=boondox]{mathalfa}

\providecommand{\thmref}[1]{Theorem~\ref{#1}}

\providecommand{\defnref}[1]{Definition~\ref{#1}}
\providecommand{\secref}[1]{Section~\ref{#1}}
\providecommand{\lemref}[1]{Lemma~\ref{#1}}

\providecommand{\figref}[1]{Figure~\ref{#1}}

\providecommand{\appref}[1]{Appendix~\ref{#1}}

\providecommand{\schemeref}[1]{Scheme~\ref{#1}}

\usepackage{refcount}

\newcommand{\eg}{e.g.}

\newcommand{\reals}{\mathbb{R}}
\newcommand{\preals}{\mathbb{R}_{\geq 0}}
\newcommand{\ints}{\mathbb{Z}}
\newcommand{\pints}{\mathbb{Z}_{\geq 0}}

\newcommand{\nats}{\mathbb{N}}

\newcommand{\ceil}[1]{\left\lceil{#1}\right\rceil}
\newcommand{\floor}[1]{\left\lfloor{#1}\right\rfloor}

\providecommand{\Exp}[1]{{\operatorname{exp} \left\{ #1 \right\}}}
\providecommand{\E}[1]{\ensuremath{\mathbb{E} \left[ #1 \right]}}

\providecommand{\h}[1]{\ensuremath{h \left( #1 \right)}}

\newcommand{\ov}{\bar{v}}

\newcommand{\oP}{\bar{P}}

\providecommand{\cE}{\mathcal{E}}

\newcommand{\Comment}[1]{}
\newcommand{\old}[1]{}
\newcommand{\rem}[1]{}

\newcommand{\eps}{{\epsilon}}

\providecommand{\hB}{{\hat{B}}}

\providecommand{\hS}{{\hat{S}}}

\providecommand{\hW}{{\hat{W}}}

\newcommand{\tE}{\tilde{E}}

\newcommand{\tr}{\tilde r}

\newcommand{\abs}[1]{\left| #1 \right|}

\providecommand{\comment}[1]{}

\newcommand{\beqn}[1]{\begin{eqnarray}\label{#1}}
\newcommand{\eeqn}{\end{eqnarray}}
\newcommand{\beq}[1]{\begin{equation}\label{#1}}
\newcommand{\eeq}{\end{equation}}

\providecommand{\var}[1]{{\mathrm{Var}\left( #1 \right)}}
\providecommand{\cov}[2]{{\mathrm{Cov}\left( #1, #2 \right)}}
\providecommand{\covSQR}[2]{{\mathrm{Cov}^2\left( #1, #2 \right)}}

\newcommand{\MSE}{\mathrm{MSE}}

\providecommand{\half}{\frac{1}{2}}

\providecommand{\half}{\frac{1}{2}}

\makeatletter
\newcommand{\vast}{\bBigg@{4}}
\newcommand{\Vast}{\bBigg@{5}}
\makeatother

\providecommand{\KL}[2]{{\bbD \left( #1 \middle\| #2 \right)}}

\providecommand{\h}[1]{{h \left( #1 \right)}}

\providecommand{\bbD}{\mathbb{D}}

\providecommand{\E}[1]{\bbE \left[ #1 \right]}

\newcommand{\PR}[1]{\Pr\left( #1 \right)}

\newcommand{\SIGsqr}[1]{\sigma^2\left( #1 \right)}

\providecommand{\Exp}[1]{\ensuremath{{\operatorname{exp} \left\{ #1
\right\}}}}

\providecommand{\BLER}[1]{P_e \left( #1 \right)}

\providecommand{\Exp}[1]{\exp \left\{ #1 \right\}}

\providecommand{\lrp}[1]{\left( #1 \right)}
\providecommand{\lrr}[1]{\left( #1 \right)}
\providecommand{\lrs}[1]{\left[ #1 \right]}
\providecommand{\lrc}[1]{\left\{ #1 \right\}}
\providecommand{\lrpm}[2]{\left( #1 \middle| #2 \right)}

\providecommand{\lrcm}[2]{\left\{ #1 \middle| #2 \right\}}

\providecommand{\IV}{\mathrm{V}}

\providecommand{\element}[3]{#1_{#2}\lrp{#3}}

\providecommand{\HB}[3]{\hB_{#1} \lrpm{#2}{#3}}
\providecommand{\M}[2]{{\element{\mathrm{MSE}}{#1}{#2}}}

\providecommand{\X}[2]{{\element{X}{#1}{#2}}}
\providecommand{\Xsqr}[2]{{\element{X^2}{#1}{#2}}}

\newcommand{\D}[1]{d_{#1}}

\providecommand{\Y}[2]{{\element{Y}{#1}{#2}}}
\providecommand{\Z}[2]{{\element{Z}{#1}{#2}}}
\providecommand{\Zeff}[2]{{\element{Z^\mathrm{eff}}{#1}{#2}}}

\providecommand{\A}[2]{{\element{{\alpha}}{#1}{#2}}}
\providecommand{\G}[2]{{\element{{\gamma}}{#1}{#2}}}
\providecommand{\B}[2]{{\element{{\beta}}{#1}{#2}}}
\providecommand{\Bsqr}[2]{\beta^2_{#1}\lrp{#2}}
\providecommand{\HS}[2]{{\element{{\hS}}{#1}{#2}}}

\providecommand{\ERR}[2]{{\element{\cE}{#1}{#2}}}
\providecommand{\ERRfin}[3]{{\element{\cE^{#1}}{#2}{#3}}}
\providecommand{\HSfin}[3]{{\element{\hS^{#1}}{#2}{#3}}}

\providecommand{\pe}[3]{\eps_{#1}\lrp{ #2 \middle| #3 }}

\providecommand{\packet}[1]{W \lrp{#1}}
\providecommand{\hpacket}[3]{\hW_{#1} \lrpm{#2}{#3}}
\providecommand{\psize}{\ell}
\providecommand{\bit}[1]{B \lrp{#1}}
\providecommand{\period}{T}
\providecommand{\size}{\ell}

%% file: figs/relays_option2.tikz
\begin{tikzpicture}[auto, arrow/.style={very thick, ->, >=stealth'},start chain=going right,>=latex,node distance=25mm,>=latex']
    \node[block] (input) {\begin{tabular}{c} Causal \\ source \end{tabular}};
    \node[block, right of = input, node distance = 45mm] (enc) {\begin{tabular}{c} Transmitter \\ (Node 0) \end{tabular}};
    \node[sum, right = 15mm of enc] (link1) {\bf \Large +};
    \node[block, right = 15mm of link1] (node2) {Node 1};
    \node[sum, right = 15mm of node2] (link2) {\bf \Large +};
    \node[mwe, right = 15mm of link2] (etc) {$\bullet\bullet\bullet$};
    \node[block, right = 15mm of etc] (dec) {\begin{tabular}{c} Node $r$ \end{tabular}};
    \node[mwe, right = 15mm of dec] (etc2) {$\bullet\bullet\bullet$};
    \node[mwe, below of = etc, node distance = 16.3mm] (FBr) {$\bullet\bullet\bullet$};
    
    \node[coordinate, above right = 20mm and 20mm of dec] (above_output) {};
    \node[coordinate, above right = 20mm and 20mm of node2] (above_node2) {};
    
    \node [above of=link1, node distance = 15mm] (noise1) {$\Z{0}{t}$};
    \node [above of=link2, node distance = 15mm] (noise2) {$\Z{1}{t}$};

    \draw[BitPipe] (input) --
    node[above, pos=0.5]{\begin{tabular}{c}
          Stream of 
       \\ packets
       \\ $\packet{\cdot}$
     \end{tabular}}
    (enc);

    \draw[BitPipe] (dec) |- 
    node [above, pos=0.75] {\begin{tabular}{c} 
                             \\ \\ 
                                Decoded packets 
                             \\ $\hpacket{r}{\cdot}{t}$ 
                          \end{tabular}}
    (above_output);

    \draw[BitPipe] (node2) |- 
    node [above, pos=0.75] {\begin{tabular}{c} 
                             \\ \\ 
                                Decoded packets 
                             \\ $\hpacket{1}{\cdot}{t}$ 
                          \end{tabular}}
    (above_node2);

    \draw[arrow] (enc) -- node {$\X{0}{t}$} (link1);
    \draw[arrow] (link1) -- node (y1) {$\Y{0}{t}$} (node2);
    \draw[arrow] (node2) -- node {$\X{1}{t}$} (link2);
    \draw[arrow] (link2) -- node (y2) {$\Y{1}{t}$} (etc);
    \draw[arrow] (etc) -- node (yR) {$\Y{r-1}{t}$} (dec);
    \draw[arrow] (dec) -- node (yRp1) {$\X{r}{t}$} (etc2);

    \draw[arrow] (noise1) -- (link1);
    \draw[arrow] (noise2) -- (link2);

    \draw[arrow] (y1) to++(0,-20mm)-| node [below,pos=.25] {} (enc);
    \draw[arrow] (y2) to++(0,-20mm)-| node [below,pos=.25] {} (node2);
    \draw[very thick] (yR) |- (FBr);
\end{tikzpicture}

%% file: main.bbl
\begin{thebibliography}{10}
\providecommand{\url}[1]{#1}
\csname url@samestyle\endcsname
\providecommand{\newblock}{\relax}
\providecommand{\bibinfo}[2]{#2}
\providecommand{\BIBentrySTDinterwordspacing}{\spaceskip=0pt\relax}
\providecommand{\BIBentryALTinterwordstretchfactor}{4}
\providecommand{\BIBentryALTinterwordspacing}{\spaceskip=\fontdimen2\font plus
\BIBentryALTinterwordstretchfactor\fontdimen3\font minus \fontdimen4\font\relax}
\providecommand{\BIBforeignlanguage}[2]{{%
\expandafter\ifx\csname l@#1\endcsname\relax
\typeout{** WARNING: IEEEtran.bst: No hyphenation pattern has been}%
\typeout{** loaded for the language `#1'. Using the pattern for}%
\typeout{** the default language instead.}%
\else
\language=\csname l@#1\endcsname
\fi
#2}}
\providecommand{\BIBdecl}{\relax}
\BIBdecl

\bibitem{Waxman-Kocsis-Stys:Axons:Book1995}
S.~G. Waxman, J.~D. Kocsis, and P.~K. Stys, \emph{The Axon: Structure, Function, and Pathophysiology}.\hskip 1em plus 0.5em minus 0.4em\relax Oxford, UK: Oxford University Press, 1995.

\bibitem{Girault-Peles:nodes-of-Ranvier:OpNeuroBio2002}
J.-A. Girault and E.~Peles, ``Development of nodes of {Ranvier},'' \emph{Current opinion in neurobiology}, vol.~12, no.~5, pp. 476--485, 2002.

\bibitem{Berger-Levy:neural-comm:TIT2010}
T.~Berger and W.~B. Levy, ``A mathematical theory of energy efficient neural computation and communication,'' vol.~56, no.~2, pp. 852--874, 2010.

\bibitem{Suksompong-Berger:integrate-and-fire-capacity:TIT2010}
P.~Suksompong and T.~Berger, ``Capacity analysis for integrate-and-fire neurons with descending action potential thresholds,'' \emph{IEEE Transactions on Information Theory}, vol.~56, no.~2, pp. 838--851, 2010.

\bibitem{Heinovski-Dressler:platooning:VNC2018}
J.~Heinovski and F.~Dressler, ``Platoon formation: Optimized car to platoon assignment strategies and protocols,'' in \emph{IEEE Vehicular Networking Conference (VNC)}, 2018.

\bibitem{Nawaz:platooning:TranVehic2019}
T.~Nawaz, M.~Seminara, S.~Caputo, L.~Mucchi, F.~S. Cataliotti, and J.~Catani, ``{IEEE 802.15. 7-compliant ultra-low latency relaying VLC system for safety-critical ITS},'' \emph{IEEE Transactions on Vehicular Technology}, vol.~68, no.~12, pp. 12\,040--12\,051, 2019.

\bibitem{Huleihel-Polyanskiy-Shayevitz:real-time-relaying:ISIT2019}
W.~Huleihel, Y.~Polyanskiy, and O.~Shayevitz, ``Relaying one bit across a tandem of binary-symmetric channels,'' in \emph{Proceedings of the IEEE International Symposium on Information Theory (ISIT)}, Paris, France, 2019, pp. 2928--2932.

\bibitem{Jog-Loh:Real-Time-Relaying:IT2020}
V.~Jog and P.-L. Loh, ``Teaching and learning in uncertainty,'' \emph{IEEE Transactions on Information Theory}, vol.~67, no.~1, pp. 598--615, 2020.

\bibitem{Ling-Scarlett:tandem-channel:EE:1-bit:TIT2023}
Y.~H. Ling and J.~Scarlett, ``Optimal rates of teaching and learning under uncertainty,'' \emph{IEEE Transactions on Information Theory}, vol.~67, no.~11, pp. 7067--7080, August 2021.

\bibitem{Ling-Scarlett:tandem-channel:EE:R=0:TIT2023}
------, ``Multi-bit relaying over a tandem of channels,'' \emph{IEEE Transactions on Information Theory}, vol.~69, no.~6, pp. 3511--3524, June 2023.

\bibitem{Rajagopalan-Schulman:FOCS:Real-Time-Networks:STOC1994}
S.~Rajagopalan and L.~Schulman, ``A coding theorem for distributed computation,'' in \emph{Proceedings of the ACM Symposium on Theory of Computing (STOC)}, 1994, pp. 790--799.

\bibitem{Information-Velocity:Wireless:WiOpt2015}
S.~K. Iyer and R.~Vaze, ``Achieving non-zero information velocity in wireless networks,'' in \emph{Proceedings of the Symposium on Modeling and Optimization in Mobile, Ad hoc, and Wireless Networks (WiOpt)}, 2015, pp. 584--590.

\bibitem{Ling-Scarlett:information-velocity::TIT2022}
Y.~H. Ling and J.~Scarlett, ``Simple coding techniques for many-hop relaying,'' \emph{IEEE Transactions on Information Theory}, vol.~68, no.~11, pp. 7043--7053, June 2022.

\bibitem{Fidler2010SurveyOfDetAndStochService}
M.~Fidler, ``Survey of deterministic and stochastic service curve models in the network calculus,'' \emph{IEEE Communications Surveys Tutorials}, vol.~12, no.~1, pp. 59--86, 2010.

\bibitem{Fidler2015aGuideToTheStichNetCalc}
M.~Fidler and A.~Rizk, ``A guide to the stochastic network calculus,'' \emph{IEEE Communications Surveys Tutorials}, vol.~17, no.~1, pp. 92--105, 2015.

\bibitem{chang2000performance}
C.-S. Chang, \emph{Performance Guarantees in Communication Networks}.\hskip 1em plus 0.5em minus 0.4em\relax Springer Science \& Business Media, 2000.

\bibitem{Fidler2006endtoendProb}
M.~Fidler, ``An end-to-end probabilistic network calculus with moment generating functions,'' in \emph{IEEE International Workshop on Quality of Service (IWQoS)}, 2006, pp. 261--270.

\bibitem{InformationVelocity:ErasuresWithFeedback:INFOCOM2022}
\MyStudent{E.~Domanovitz}, T.~Philosof, and \khina, ``The information velocity of packet-erasure links,'' in \emph{Proceedings of the IEEE International Conference on Computer Communications (INFOCOM)}, May 2022.

\bibitem{inovan-telatar:private-com}
R.~Inovan and E.~Telatar, private communication.

\bibitem{CausalPM:ISIT2019}
A.~Lalitha, \khina, T.~Javidi, and V.~Kostina, ``Real-time binary posterior matching,'' in \emph{Proceedings of the IEEE International Symposium on Information Theory (ISIT)}, Paris, France, Jul. 2019, pp. 2239--2243.

\bibitem{SchalkwijkKailath}
J.~P.~M. Schalkwijk and T.~Kailath, ``A coding scheme for additive noise channels with feedback--{I}: No bandwidth constraint,'' \emph{IEEE Transactions on Information Theory}, vol.~12, pp. 172--182, Apr. 1966.

\bibitem{Schalkwijk:feedback:finite-BW:TIT1966}
J.~P.~M. Schalkwijk, ``A coding scheme for additive noise channels with feedback--ii: Band-limited signals,'' \emph{IEEE Transactions on Information Theory}, vol.~12, no.~2, pp. 183--189, April 1966.

\bibitem{Gallager-Nakiboglu:SK-via-Elias:TIT2009}
R.~G. Gallager and B.~Nakibo{\u{g}}lu, ``Variations on a theme by {Schalkwijk} and {Kailath},'' vol.~56, no.~1, pp. 6--17, Dec. 2009.

\bibitem{Elias57:JSCC:BW-expansion}
P.~Elias, ``Channel capacity without coding,'' in \emph{Proceedings of the IRE}, vol.~45, no.~3, Jan. 1957, pp. 381--381.

\bibitem{ScheinGallager00}
B.~Schein and R.~G. Gallager, ``The {G}aussian parallel relay channel,'' in \emph{Proc. Int. Symp. Info. Theory (ISIT)}, Sorrento, Italy, June 2000, p.~22.

\bibitem{Laneman-Tse-Worenll:cooperative-diversity:TIT2004}
J.~N. Laneman, D.~N. Tse, and G.~W. Wornell, ``Cooperative diversity in wireless networks: Efficient protocols and outage behavior,'' \emph{IEEE Transactions on Information theory}, vol.~50, no.~12, pp. 3062--3080, Novmber 2004.

\bibitem{RematchAndForward_Full}
Y.~Kochman, \khina, \Adviser{U.~Erez}, and R.~Zamir, ``Rematch-and-forward: Joint source/channel coding for parallel relaying with spectral mismatch,'' \emph{IEEE Transactions on Information Theory}, vol.~60, no.~1, pp. 605--622, Jan. 2014.

\bibitem{Yamamoto80}
H.~Yamamoto and K.~Itoh, ``Source coding theory for multiterminal communication systems with a remote source,'' \emph{Tran.\ IECE of Japan}, vol. E\ 63, no.~10, pp. 700--706, 1980.

\bibitem{BergerZhangViswanathan:CEO:IT1996}
T.~Berger, Z.~Zhang, and H.~Viswanathan, ``The {CEO} problem [multiterminal source coding],'' \emph{IEEE Transactions on Information Theory}, vol.~42, no.~3, pp. 887--902, 1996.

\bibitem{Wagner-Tavilar-Viswanath:Gaussian-lossy-Sepian--Wolf:TITT2008}
A.~B. Wagner, S.~Tavildar, and P.~Viswanath, ``Rate region of the quadratic {Gaussian} two-encoder source-coding problem,'' \emph{IEEE Transactions on Information Theory}, vol.~54, no.~5, pp. 1938--1961, 2008.

\bibitem{Draper-Wornell:Sequential-CEO:ISIT2002}
S.~C. Draper and G.~W. Wornell, ``Successively structured {CEO} problems,'' Lausanne, Switzerland, July 2002, p.~65.

\bibitem{Draper-Wornell:SideInfo-sensors:JSC2004}
------, ``Side information aware coding strategies for sensor networks,'' \emph{IEEE Journal on Selected Areas in Communications}, vol.~22, no.~6, pp. 966--976, 2004.

\bibitem{Chen-Zhang-Berger-Wicker:distributed-sensor-networks:Allerton2003}
J.~Chen, X.~Zhang, T.~Berger, and S.~Wicker, ``Rate allocation in distributed sensor network,'' vol.~41, no.~1, Monticello, IL, USA, October 2003, pp. 531--540.

\bibitem{EquitzCover91}
W.~H.~R. Equitz and T.~M. Cover, ``Successive refinement of information,'' \emph{IEEE Transactions on Information Theory}, vol.~37, no.~2, pp. 851--857, Mar. 1991.

\bibitem{LastrasBerger}
L.~Lastras and T.~Berger, ``All sources are nearly successively refinable,'' \emph{IEEE Transactions on Information Theory}, vol.~47, pp. 918--926, March 2001.

\bibitem{PapoulisBook:4thEd}
A.~Papoulis and S.~U. Pillai, \emph{Probability, Random Variables, and Stochastic Processes}, 4th~ed.\hskip 1em plus 0.5em minus 0.4em\relax Tata McGraw-Hill Education, 2002.

\bibitem{GallagerBook1968}
R.~G. Gallager, \emph{Information Theory and Reliable Communication}.\hskip 1em plus 0.5em minus 0.4em\relax New York: John Wiley \& Sons, 1968.

\bibitem{ElGamalKimBook}
A.~{El Gamal} and Y.-H. Kim, \emph{Network Information Theory}.\hskip 1em plus 0.5em minus 0.4em\relax Cambridge University Press, 2011.

\bibitem{CoverBook2Edition}
T.~M. Cover and J.~A. Thomas, \emph{Elements of Information Theory}, 2nd~ed.\hskip 1em plus 0.5em minus 0.4em\relax New York: Wiley, 2006.

\bibitem{BenYishaiShayevitz:FeedbackWithWZ:ISIT}
A.~Ben-Yishai and O.~Shayevitz, ``The {Gaussian} channel with noisy feedback: Improving reliability via interaction,'' in \emph{Proceedings of the IEEE International Symposium on Information Theory (ISIT)}, Hong Kong, June 2015, pp. 2500--2504.

\bibitem{Wainwright:high-dim-stat:Book2019}
M.~J. Wainwright, \emph{High-Dimensional Statistics: A Non-Asymptotic V iewpoint}.\hskip 1em plus 0.5em minus 0.4em\relax Cambridge, UK: Cambridge university press, 2019, vol.~48.

\bibitem{Vershynin:high-dim-prob:Book2018}
R.~Vershynin, \emph{High-Dimensional Probability: An Introduction with Applications in Data Science}.\hskip 1em plus 0.5em minus 0.4em\relax Cambridge, UK: Cambridge university press, 2018, vol.~47.

\end{thebibliography}
